\documentclass[modern,useAMS,usenatbib,usegraphicx,twocolumn]{aastex62}

\newcommand\msun{\ensuremath{\rm M_{\odot}}}

\submitjournal{ApJ}
\usepackage{placeins}
\shorttitle{lightcurves of IIP SNe}
\shortauthors{Ricks and Dwarkadas}

\begin{document}

\title{Excavating the Explosion and Progenitor Properties of Type IIP
  Supernovae via Modelling of their Optical Lightcurves}

\correspondingauthor{Vikram Dwarkadas}
\email{vikram@astro.uchicago.edu}

\author[0000-0002-0786-7307]{Wilson Ricks}
\affil{Dept.~of Astronomy and Astrophysics \\
University of Chicago \\
5640 S Ellis Ave., Chicago, IL 60637}

\author[0000-0002-4661-7001]{Vikram V. Dwarkadas}
\affil{Dept.~of Astronomy and Astrophysics \\
University of Chicago \\
5640 S Ellis Ave., Chicago, IL 60637}



\begin{abstract}
The progenitors of Type IIP supernovae (SNe) are known to be red
supergiants, but their properties are not well determined.  We employ
hydrodynamical modelling to investigate the explosion characteristics
of eight Type IIP supernovae, and the properties of their progenitor
stars. We create evolutionary models using the {\sc MESA} stellar
evolution code, explode these models, and simulate the optical
lightcurves using the {\sc STELLA} code.  We fit the optical
lightcurves, Fe II 5169\AA\ velocity, and photospheric velocity, to
the observational data. Recent research has suggested that the
progenitors of Type IIP SNe have a zero age main sequence (ZAMS) mass
not exceeding $\sim18$ $\msun$. Our fits give a progenitor ZAMS mass
$\leq18$ $\msun$ for seven of the supernovae. Where previous
progenitor mass estimates exist, from various sources such as
hydrodynamical modelling, multi-wavelength observations, or
semi-analytic calculations, our modelling generally tends towards the
lower mass values. This result is in contrast to results from previous
hydrodynamical modelling, but is consistent with those obtained using
general-relativistic radiation-hydrodynamical codes.  We do find that
one event, SN 2015ba, has a progenitor whose mass is closer to 24
$\msun$, although we are unable to fit it well. We also derive the
amount of $^{56}$Ni required to reproduce the tail of the lightcurve,
and find values generally larger than previous estimates. Overall, we
find that it is difficult to characterize the explosion by a single
parameter, and that a range of parameters is needed.

\end{abstract}



\section{Introduction} 
\label{sec:intro}
 
Supernovae (SNe) are divided into various types based on their spectra
and sometimes their lightcurves \citep{turatto03}. SNe of Type II show
H lines in their spectrum, and are thought to be all core-collapse,
i.e. those which arise from the explosion of massive stars $\ga$ 8
$\msun$. SNe of Type I do not show H lines in their spectrum. Of the
SNe that comprise this category, Type Ia's are presumed to arise from
white dwarfs in binary systems \citep{hkrr13}. SNe of Type Ib and Ic
do not show H (Ib), and He (Ic) lines in their spectra. The lack of H
and He envelopes had early on pointed to high-mass Wolf-Rayet (W-R)
star progenitors \citep{gaskelletal86}, which are stripped of their H
and He envelopes. It is also possible that they may arise from
somewhat lower mass-stars in a binary system, where the companion star
is responsible for mass being stripped off the progenitor. In either
case it is clear that they arise from the core-collapse of a massive
star.

Type IIP SNe, which show a plateau in their optical lightcurves, are
the most common type of core-collapse SN \citep{smarttetal09,
  eldridgeetal13}. Observations show that they comprise almost 50\% of
total core-collapse SNe. The progenitors of Type IIP SNe are well
established as being red supergiant (RSG) stars \citep[see for
  example][]{smartt09}. Red supergiants are post-main-sequence massive
stars that have finished burning H and are in the He burning
phase. Stars up to about 30 $\msun$ will end their lives as RSG's,
giving rise to IIP or perhaps IIL (which show a linear drop in the
lightcurve, in contrast to the plateau seen in the IIPs) SNe
\citep{ekstrometal13}. Stars with initial mass above 30 $\msun$, but
lower than about 40 $\msun$, may pass through a RSG phase but end
their lives as Wolf-Rayet stars. The range may vary somewhat depending
on the actual mass-loss rates of these stars throughout their
evolution, and factors like rotation and magnetic fields, which are
only recently being taken into consideration.

The fates of massive stars, and the progenitors of the various types
of SNe, are an active area of research \citep{galyametal07}. Even
though several thousand SNe are known, the relationship between the
massive stars that core-collapse and the type of SNe that they go on
to form is not well established. In this context, it may seem that the
IIPs are in a special position, as their progenitors are
clearly red supergiants, and the SNe themselves are visibly
identifiable via the plateau in their lightcurves. But the parameters
that determine the properties of the lightcurve, such as its shape,
the duration and luminosity of the plateau, and the emission beyond
the plateau, their relation to the stellar parameters, especially the
zero age main sequence (ZAMS) stellar mass, and the SN explosion
parameters, such as the explosion energy and $^{56}$Ni mass, are not
well understood \citep{faranetal14, nakaretal16}.

Early observations of SN progenitors suggested progenitor masses of
Type IIP's generally below 20 $\msun$ \citep{Lietal06, hendryetal06,
  lietal07}, leading to suggestions that Type IIP SNe arose from
progenitors $<$ 20 $\msun$ \citep{lietal07}. This was better
quantified by \citet{smartt09} who found, in their study of optically
identified Type IIP SN progenitors, that IIP progenitors did not seem
to exceed 16.5 $ \pm$ 1.5 $\msun$. This has come to be known as the
red supergiant problem. \citet{yc10} found that pulsationally driven
super winds could change the evolution of a star of initial mass $> 19
\msun$, causing it to become a Ib or IIb SN, or perhaps even a IIn,
but no longer a IIp. \citet{ekstrometal12} suggested another
possibility, that the outer layers could exceed the Eddington limit,
resulting in enhanced mass-loss. Whatever the reason,
\citet{georgyetal12} showed that with an enhanced mass-loss rate,
rotating stars above 16.8 $\msun$, and non-rotating stars above 19
$\msun$, would end their lives as W-R stars rather than as RSGs, and
would not give rise to IIP SNe. \citet{horiuchietal14} and
\citet{kochanek15} suggested that the RSG problem may be understood if
stars above a certain mass limit collapse directly to black holes, and
that the value of the compactness parameter may determine the boundary
between successful and failed explosions. There have also been
suggestions that dust extinction \citep{we12}, or the increasing
bolometric correction \citep{db18}, have not been properly taken into
account and could help to mitigate the problem.

\citet{vvd14} showed, from an analysis of X-ray lightcurves of SNe,
that the lack of X-ray bright Type IIPs seemed to indicate an upper
mass-loss limit, and thereby an upper mass limit, of about 19 $\msun$
for IIP progenitors.  \citet{smartt15}, using a larger sample,
reassessed the optical data and concluded that the problem was even
more severe, that observed populations of supernovae in the local
universe are not produced by stars $>$ 18 \msun, and that most stars
with initial masses above this value would collapse directly to black
holes without leaving a visible SN. The latter assumption has also
received some support from theoretical calculations of stellar
collapse by \citet{sukhboldetal16}, who have found that only
  about 10\% of SNe arise from stars with ZAMS masses $> 20
  \msun$. Observations of a RSG in NGC 6946 that faded away over a
decade, with no indications of a SN explosion or debris
\citep{adamsetal17}, have provided further impetus to this line of
reasoning.

A very large dataset of {\em confirmed} progenitor masses of SNe is
needed to decipher how massive stars end their lives, if (and whether)
they core-collapse to a SN or go directly to black holes, and what
type of SN results from this. Unfortunately, confirming progenitor
masses is a very difficult task. Several efforts are underway to
optically detect SN progenitors \citep{smarttetal09, smartt15, er16,
  maund17, vandyk17}.  While direct optical identification of
progenitors continues, it is a slow and time-consuming process that
depends on the availability of high resolution imaging in the past.

There also exist indirect ways of learning more about the SN explosion
and progenitor. One of these is by modeling the optical lightcurves of
the SN starting from the evolution of the pre-SN star. Type IIP SNe
are characterized by a distinct and long-lasting (several months)
plateau in their optical lightcurve. Parameters involved in simulating
the lightcurve, such as the initial rise, plateau luminosity, duration
of the plateau, and slope of the tail, can provide information on
various explosion parameters such as the { $^{56}$Ni} mass,
explosion energy, and the presence of circumstellar material around
the SN. The evolution of the star, its collapse to form a SN, and the
accurate modelling of the lightcurve, can provide a measure of the
initial ZAMS mass. Hitherto, this has always been a time-consuming and
computationally expensive endeavor, requiring several steps:
\begin{enumerate}
\item Modelling the evolution of the high mass star (that gave rise to the
SN) up to core collapse, using a stellar evolution code.
\item Modelling the explosion of the star to give rise to a SN. 
\item Using a radiation hydrodynamics code to model the SN lightcurve.
\item Reiterating steps 1-3 till a good model fit is obtained.
\end{enumerate}
\noindent
Recent advances have however made such calculations more feasible. The
release and continued development of the {\sc MESA} code
\citep{mesa1,mesa2,mesa3,mesa4} has provided astronomers with access
to a modern stellar evolution code that includes a variety of physics,
and the ability to construct models for stars of most initial masses
and metallicities, taking various mass-loss prescriptions into
account. The {\sc SNEC} code \citep{snecpaper} was made available
to model the explosion of SNe and calculate the resulting lightcurve
in various bands. The combination of {\sc MESA} and
{\sc SNEC} has been used by several authors
\citep{morozovaetal16, dr17, patnaudeetal17}. More recently,
\citet{mesa4} have provided a complete recipe to accomplish this task
by combining the {\sc MESA} code with a reduced version of the
{\sc STELLA} code \citep{blinnikovetal98, bs04,
  blinnikovetal06}, which allows for all of the steps necessary in
calculating the lightcurves to be completed entirely within the
{\sc MESA} framework. This development allows for the entire
process, from the initiation of the stellar model to the production of
the optical lightcurve, to be accomplished in less than a
day. As is to be expected, MESA cannot deal with each step in
  all its complexity. In particular it does not attempt to model the
  micro-physics of the SN explosion itself, which would be a huge
  task. Instead, the model star is exploded through a mechanism that
  artificially imparts the required energy and some other parameters,
  thus leaving some freedom in how these are calculated. We note that
  this is not unusual - a similar technique is utilized in the SNEC
  code for example. The method used in {\sc MESA} is further
  described in \S 2.

In this paper our goal is to explore the properties of recent Type IIP
SN explosions, to evaluate both the explosion characteristics as well
as the properties of the exploded star, to study the relationships, if
any, between the factors that determine the shape and luminosity of
the IIp lightcurve and the properties of the SN explosion, and to
unearth the progenitor mass. In order to accomplish this, we use the
combination of {\sc MESA} and {\sc STELLA} codes to simulate the
lightcurves of several Type IIP SNe, and compare to the
observations. A good fit to both the lightcurves and photospheric
velocities allows us to constrain the explosion and stellar properties
and to thereby determine the progenitor mass. In \S \ref{sec:mesa} we
outline the basic procedure used in calculating the lightcurves with
{\sc MESA} and {\sc STELLA}. \S 3 displays the application of this
technique to match the observed lightcurves, and photospheric
velocities, for a set of Type IIP SNe. Each SN is discussed in
detail. \S 4 displays the relationships between various parameters,
including stellar mass, { $^{56}$Ni} mass, and explosion energy
from our work, and compares them to those in the literature. Finally,
\S 5 summarizes our research, discusses further prospects, and
revisits the conclusions for the progenitors of Type IIP SNe.

\section{ Using MESA and STELLA to compute lightcurves}
\label{sec:mesa}

{\sc MESA} is a state-of-the-art, one-dimensional, modular,
open-source suite for stellar evolution
\citep{mesa1,mesa2,mesa3,mesa4}. A variety of physics modules are
included, which allow for modelling of single stars as well as those
in binary systems. The suite of tools has recently been extended
\citep{mesa4} to include the explosion of massive stars and the
modelling of SN lightcurves. In the present work, done using the 10398
release of {\sc MESA}, the code is used to model the evolution of high
mass stars ($>$ 8 $\msun$) from the proto-stellar phase until they
form an Fe core, at which point they eventually core collapse as
SNe. {\sc MESA} then simulates the SN explosion by removing the
proto-neutron star, allowing the model to continue infall until its
inner boundary reaches 200 km, and then injecting a specified amount
of energy into a thin layer near this inner boundary to induce the
explosion. {Because the explosion is not explicitly modeled, the user
  must also specify the total mass of synthesized $^{56}$Ni, whose
  value is adjusted in our calculations such that the simulated light
  curves match the observed ones. {\sc MESA} handles the SN shock
  propagation until just before breakout, at which point {\sc STELLA}
  takes over.

{\sc STELLA} models the breakout itself and the shock interaction with
circumstellar material through the nebular phase. It computes the
primary SN observables over this period. The limited version of {\sc
  STELLA} packaged with {\sc MESA} includes 40-200 frequency groups,
which is adequate to produce lightcurves, but not SN
spectra. Therefore in this paper we have chosen to compare the models
to observations of the lightcurves. {\sc STELLA} provides not only the
total bolometric lightcurve, but also lightcurves in the U, B, V, R,
and I passbands \citep{bessell05}.  As pointed out in \citet{mesa4},
comparing the simulated lightcurves with the observational ones using
the UBVRI filters, and the resulting (quasi)-bolometric lightcurves,
can result in a degeneracy in the progenitor mass. This degeneracy can
in many cases be removed by modelling the photospheric velocity in
addition to the optical lightcurve (although see
  \cite{goldberg19}). Since the photospheric velocity itself is not
  observable, the Fe II 5169\AA\,velocity is used as a proxy for the
  photospheric velocity. The photospheric velocity is generally
  calculated at an optical depth of $\tau = 2/3$, whereas the velocity
  of the Fe II line is calculated in the Sobolev approximation, using
  a {\it Sobolev} optical depth $\tau_{Sob} = 1$. The Sobolev
  approximation, and the resulting value of $\tau_{Sob}$ used in MESA
  is valid in so far as the ejecta are expanding
  homologously. Homologous expansion however is not reached until
  roughly 20-30 days post explosion \citep{mesa4}. Therefore the Fe II
  velocities are not calculated de facto in the MESA code prior to 25
  days. While it is possible to calculate them at earlier times, the
  numbers are unphysical and invalid if expansion is not
  homologous. In order to compare to the observed Fe II 5169
  \AA\ velocity at these early times, we therefore use the
  photospheric velocity calculated at an optical depth $\tau =2/3$. In
  our plots we therefore show both calculated velocities: the Fe II
  5169 \AA\ velocity from day 25 onwards, and the photospheric
  velocity from the time of explosion. Although the Fe II velocities
  tend to be higher than the photospheric velocities during the
  plateau phase, \citet{mesa4} note that there should be little
  difference between the two at early times. In summary, for each SN,
  we model the quasi-bolometric lightcurve, the lightcurves in the
  UBVRI Johnson filters (or any of these that are provided) the Fe II
  5169\AA\,velocity, and the photospheric velocity.

 We use the {\sc MESA} test suite inlists \verb|example_make_pre_ccsn|
 and \verb|example_ccsn_IIp| to model the evolution of the star until
 the core-collapse SN phase. The main parameters that we vary are the
 ZAMS mass, mass-loss efficiency and the rotation velocity. Unless
 otherwise specified, all models referenced in this paper assume solar
 metallicity. For all models, overshooting and mixing length
 parameters are the same as given in the {\sc MESA} defaults: {\it
   f}$_{\rm{ov}}=0.01$}, {\it f}$_{0,\rm{ov}}=0.004$, and
$\alpha_{\rm{MLT}}=3.0$.

\begin{figure}[!htb]
\includegraphics[width=\columnwidth]{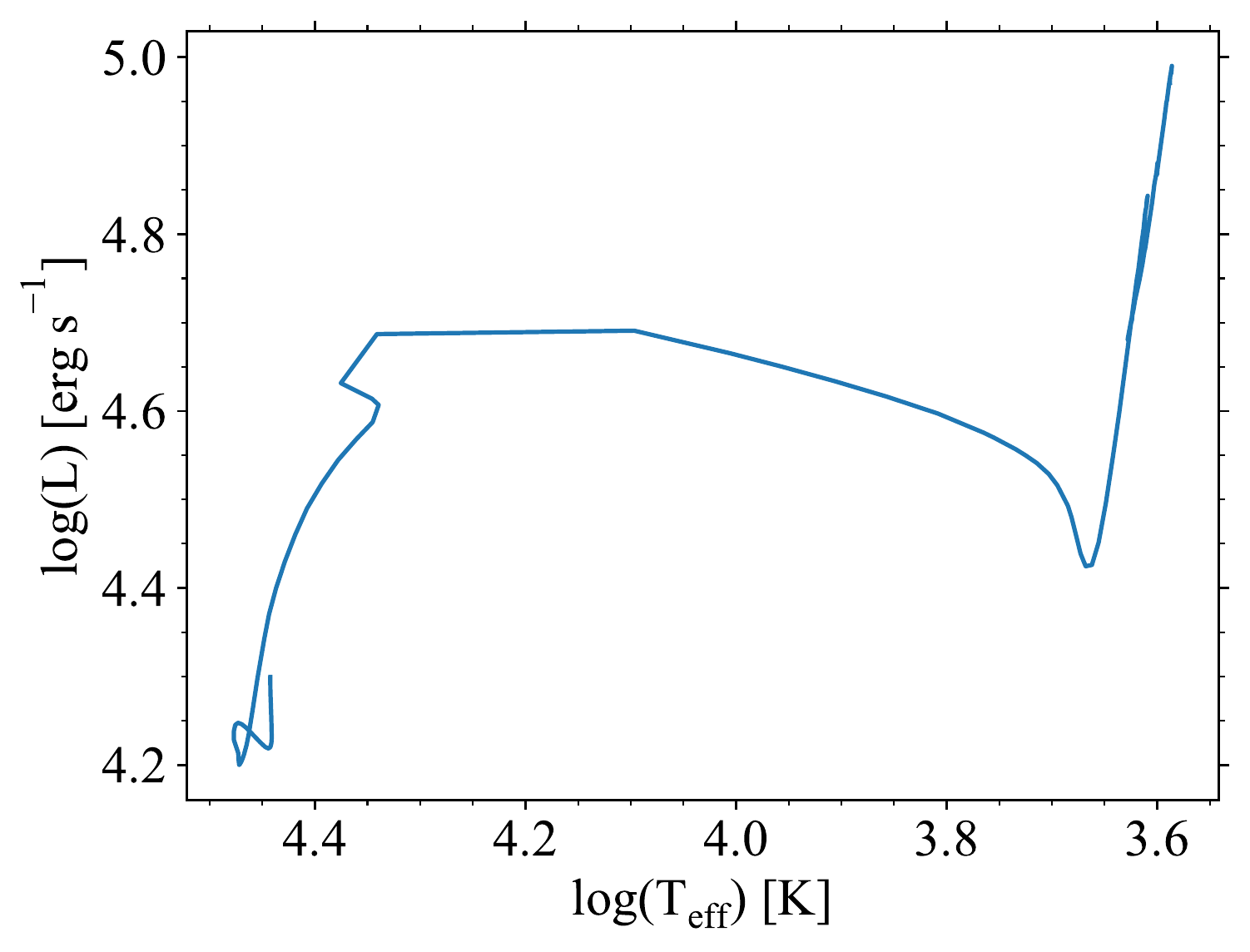}
\caption{The evolutionary HR diagram for the best-fit {\sc MESA} progenitor model
  corresponding to SN 2014cx.
\label{fig:hr2014cx}}
\end{figure}

As an example of the stellar evolution modelling, in Figure
\ref{fig:hr2014cx} we show the HR diagram for the evolution of the
progenitor star of SN 2014cx. From our lightcurve modelling, the best
fit was produced for a progenitor ZAMS star of mass 12 $\msun$. The
evolutionary track found from the model is as expected for the
evolution of a 12 $\msun$ star (see for instance
\citet{ekstrometal12}) which spends most of its life in the main
sequence and ends its life as a red supergiant, giving rise to a Type
IIP SN. The parameters of this star can be found in Table 1.

\begin{figure}[!htb]
\includegraphics[width=\columnwidth]{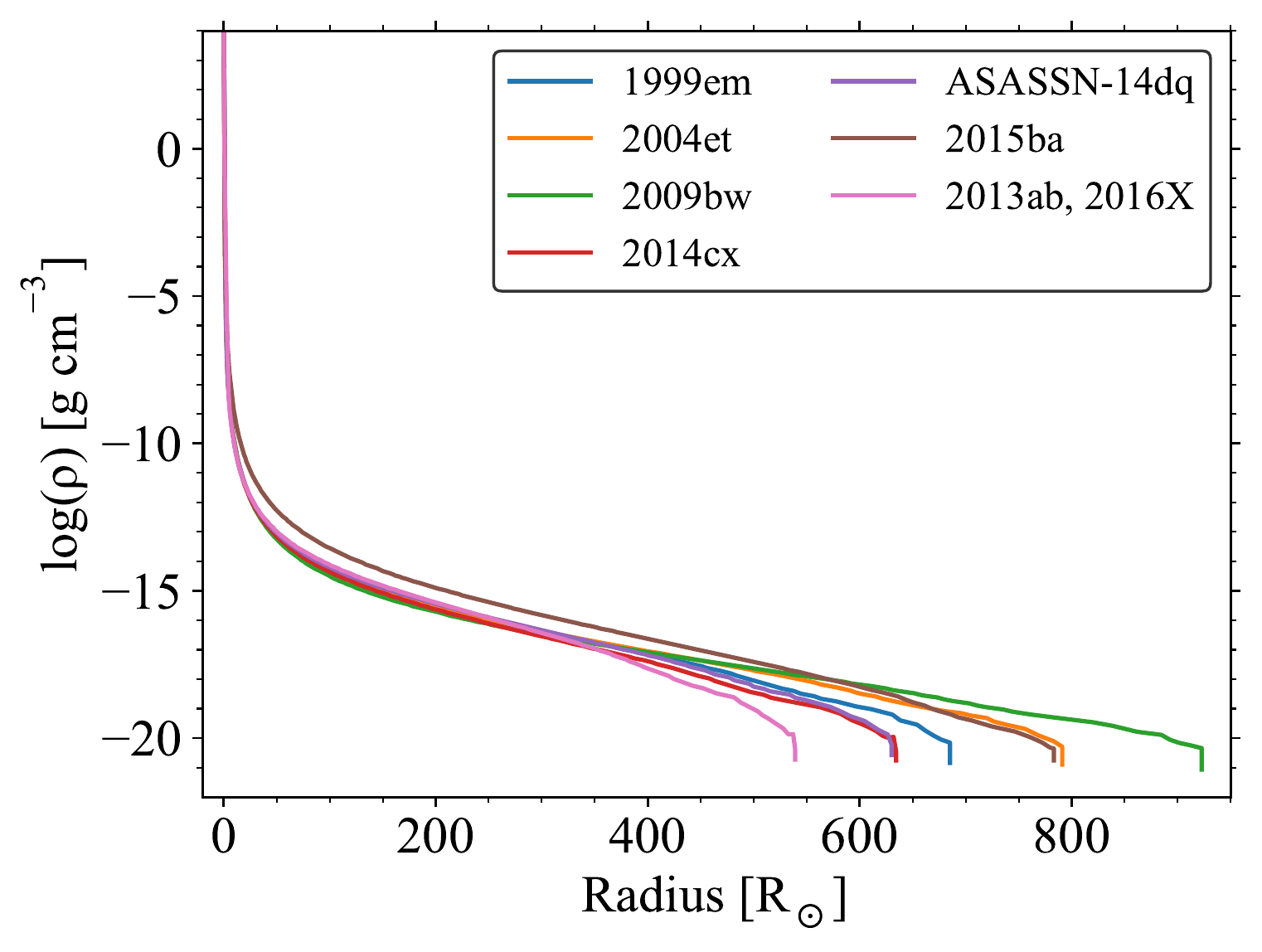}
\caption{The density distributions of the best-fit MESA progenitor
  models for all eight SNe investigated in this paper, just before
  {explosion}. \label{fig:general}
\label{fig:dens}}
\end{figure}

In Figure \ref{fig:dens} we show the final density distributions (just
before explosion) for the eight SN progenitor models computed in this
paper. {\sc MESA} allows a specified amount of circumstellar material
(CSM) around the SN to be included, and we have found that doing so
generally produces a better fit at early epochs, as was noted by
\citet{morozova18}. This material, added by {\sc MESA} just before the
handoff to {\sc STELLA} and presumably ejected in literally the last
couple of years of the star's life, has been found necessary to fit
the initial light curves.

Once the stellar evolution model is completed, the SN is allowed to
explode. Parameters for the explosion itself, such as total energy
injection, and total { $^{56}$Ni} mass, are adjusted till a good
match between the simulated and observed lightcurves is obtained.  The
lightcurves and Fe II velocities computed by {\sc STELLA} are compared
to the observations, and the model is refined until a suitable match
is found.  Although parameters for SNe that have been modelled by
other authors are given in the literature, there is considerable
variation between these, and we do not regard these as a viable
starting point. The effect of varying some parameters can be found in
\citet{mesa4}, although the cumulative effect of varying several
parameters at a time can be more complex.

The number of variables involved in generating the lightcurves is
extremely large. Tables 1 and 2 list the major parameters that were
varied, but there may be cases where other stellar parameters or
explosion parameters may be needed. While one method of fitting
observed and simulated lightcurves is to generate a large grid of
lightcurves encompassing the entire range of values, this is not
computationally feasible given the enormous range. Instead, we have
chosen to use a combination of science, brute force, and our own
experience in fitting the lightcurves, with the help of the parameter
variations shown in \citet{mesa4} and a preliminary study. Inspecting
the lightcurves and Fe II velocity, we decide on a range of initial
values to use, and then continually refine the parameters until what
we deem is a reasonable fit is obtained. Often, as will be seen, the
decision is not so clear, because a single parameter may make the
bolometric or UBVRI lightcurves better but the velocity worse, or vice
versa, leaving us to determine which one should be given more weight,
or find a compromise. Given our eyeballing technique, we have not
attempted to make any quantitative measurements of the goodness of
fit.

\citet[][hereafter M18]{morozova18} have modelled the lightcurves of
Type IIP SNe using the {\sc KEPLER} stellar evolution code combined
with the {\sc SNEC} radiation hydrodynamics code. They have used a
two-step fitting method employing a restricted grid of parameters,
which they compare to the lightcurves. They did not compute and
compare to photospheric velocities. For those SNe that are common
between the two papers, we have provided a comparison between our
results and theirs.

\section{Light-curve Fits for SNe}
\label{section:fits}
As mentioned above, our standard technique involves fitting the
quasi-bolometric lightcurve (a bolometric lightcurve derived from the
individual U, B, V, R, and I color curves), the U, B, V, R, and I
color curves themselves, and the FeII 5169 \AA\,velocity. In this
section we present the results of modelling the lightcurves of a set
of SNe. After presenting the best-fit models for each SN, we compare
our findings to prior results.

In total we have considered eight SNe. Two of these, SN 1999em and SN
2004et, were chosen because they have been well studied in the past,
and allow us to compare our simulated parameters with those reported
in the literature. We use these to verify that the {\sc MESA} and {\sc
  STELLA} computations provide reasonable results that fall within the
range of acceptable values. It also allows us to check the agreement
between the derived parameters, and those obtained via other means
such as optical identification of progenitors, theoretical and
multiwavelength modelling, or hydrodynamical simulations.

Having assessed the validity of the {\sc MESA} results, we proceed to
tackle six more SNe which have not been widely studied in the
literature. Some of these SNe have been reported as having progenitor
masses in excess of 18 $\msun$, such as SNe 2014cx, ASASSN-14dq,
2015ba, and 2016X. For each SN, we present fits to the UBVRI
quasi-bolometric lightcurve, individual U, B, V, R and I color curves,
and Fe II 5169 velocities. Our set of SNe is constrained by the
requirement that all of this observational data be available to
us. For many SNe, this was not the case as no tables were provided.
Often the photospheric velocities were missing, as well as one or more
of the necessary color curves. We made a minor exception in the case
of SN 2015ba, a SN with a very interesting light curve that we decided
to include despite its lack of data in the U, R and I bands. We have
also tried to include a variety of SNe, and discarded those with
similar lightcurves and photospheric velocities to one already in our
set. All quasi-bolometric lightcurve and Fe II 5169 observational data
referenced throughout this paper have been digitized from available
figures. Our {\sc STELLA} runs used 120 frequency bins rather than the
default 40 to better model the individual color curves.

Tables 1 and 2 summarize the parameters and properties for the
evolutionary models, and for the SN explosion, from our light-curve
fitting. Table 1 lists the model parameters, especially the zero age
main sequence mass, the rotation velocity (as a function of the
critical velocity), the metallicity, and the wind efficiency. Table 2
gives the explosion parameters, such as the mass and radius of the
star prior to explosion, the core mass, the wind properties that gave
rise to the circumstellar material, the explosion energy, the
$^{56}$Ni mass and the plateau luminosity. We note that in what
follows, we refer to the explosion energy of 10$^{51}$ erg as 1 foe,
as is often done in the literature. UT dates are used throughout this
paper.

\begin{deluxetable*}{c|cccc}[]
\centering \tablecaption{Properties and parameters of SN progenitor
  models. The ``SN'' column lists the supernova whose progenitor is
  being modelled. M$_{ZAMS}$ is the ZAMS mass of the progenitor,
  $(\nu/\nu_c)_{ZAMS}$ is the rotation velocity of the star in terms
  of the critical rotation velocity, $Z$ is the initial metallicity
  and $\eta_{\text{wind}}$ is the scaling factor for mass-loss
    efficiency during stellar evolution.  \label{tab:table}}
\tablehead{ \colhead{SN} $\;$ & \colhead{M$_{ZAMS}$ [M$_\odot$]} &
  \colhead{$(\nu/\nu_c)_{ZAMS}$} & \colhead{$Z$} &
  \colhead{$\eta_{\text{wind}}$}}

\startdata
1999em & 14 & 0.2 & 0.02 & 1.0\\
2004et & 16 & 0 & 0.02 & 0.8 \\
2009bw & 18 & 0 & 0.02 & 0.8 \\
2013ab & 11 & 0.35 & 0.02 & 1.0 \\
2014cx & 12 & 0.35 & 0.02 & 1.0 \\
ASASSN-14dq & 13 & 0.25 & 0.014 & 1.0 \\
2015ba & 24 & 0 & 0.02 & 0.4 \\
2016X  & 11 & 0.35 & 0.02 & 1.0 \\
\enddata
\end{deluxetable*}

\begin{deluxetable*}{c|ccccccccc}[!htbp]
\tablecaption{Properties and parameters of the SN explosion
  models. The ``SN'' column lists the supernova being
  modelled. M$_{exp}$ is the progenitor mass at the time of explosion,
  M$_{ej}$ is the ejecta mass and R$_{exp}$ is the progenitor radius
  at the time of explosion. $t_{CSM}$ is the number of years for the
  CSM wind artificially placed outside the model, $\dot{M}_{CSM}$ is
  the mass loss rate from this wind, and $v_{CSM}$ is the wind
  velocity. E$_{exp}$ is the total energy injected into the model
  during the SN explosion, while { M$_{^{56}Ni}$} is the total
  $^{56}$Ni mass. The last column gives the bolometric luminosity of
  the optical lightcurve at 50 days post-explosion. \label{tab:table}}
\centering \tablehead{ \colhead{SN} $\;$ & \colhead{M$_{exp}$} &
  \colhead{M$_{ej}$} & \colhead{R$_{exp}$} & \colhead{$t_{CSM}$ } &
  \colhead{$\dot{M}_{CSM}$ } & \colhead{$v_{CSM}$ } &
  \colhead{E$_{exp}$ } & \colhead{{ M$_{^{56}Ni}$} } &
  \colhead{$log(\text{L}_{pl}/\text{L}_\odot)$} \\ \colhead{} &
  \colhead{[M$_\odot$]} & \colhead{[M$_\odot$]} &
  \colhead{[R$_\odot$]} & \colhead{[yr]} & \colhead{[M$_\odot$
      yr$^{-1}$]} & \colhead{[km s$^{-1}$]} & \colhead{[$10^{51}$
      erg]} & \colhead{[M$_\odot$]} } \startdata 1999em & 11.83 &
10.28 & 682 & 1.4 & 0.15 & 10 & 0.55 & 0.075 & 8.51 \\ 2004et & 13.69
& 11.87 & 792 & 1.4 & 0.30 & 10 & 0.90 & 0.100 & 8.67 \\ 2009bw &
16.27 & 14.26 & 911 & 1.4 & 0.30 & 10 & 0.64 & 0.060 & 8.52 \\ 2013ab
& 9.57 & 8.07 & 536 & 1.4 & 0.15 & 10 & 0.65 & 0.065 & 8.54 \\ 2014cx
& 10.14 & 8.66 & 636 & 1.4 & 0.15 & 10 & 0.70 & 0.100 & 8.62
\\ ASASSN-14dq & 11.41 & 9.93 & 629 & 1.4 & 0.30 & 10 & 0.95 & 0.075 &
8.67 \\ 2015ba & 22.58 & 20.17 & 784 & 1.4 & 0.50 & 10 & 0.85 & 0.050
& 8.59 \\ 2016X & 9.57 & 8.07 & 536 & 1.4 & 0.30 & 10 & 0.60 & 0.036 &
8.52 \\ \enddata
\end{deluxetable*}

\subsection{SN 1999em}
\begin{figure}[!htb]
\includegraphics[width=\columnwidth]{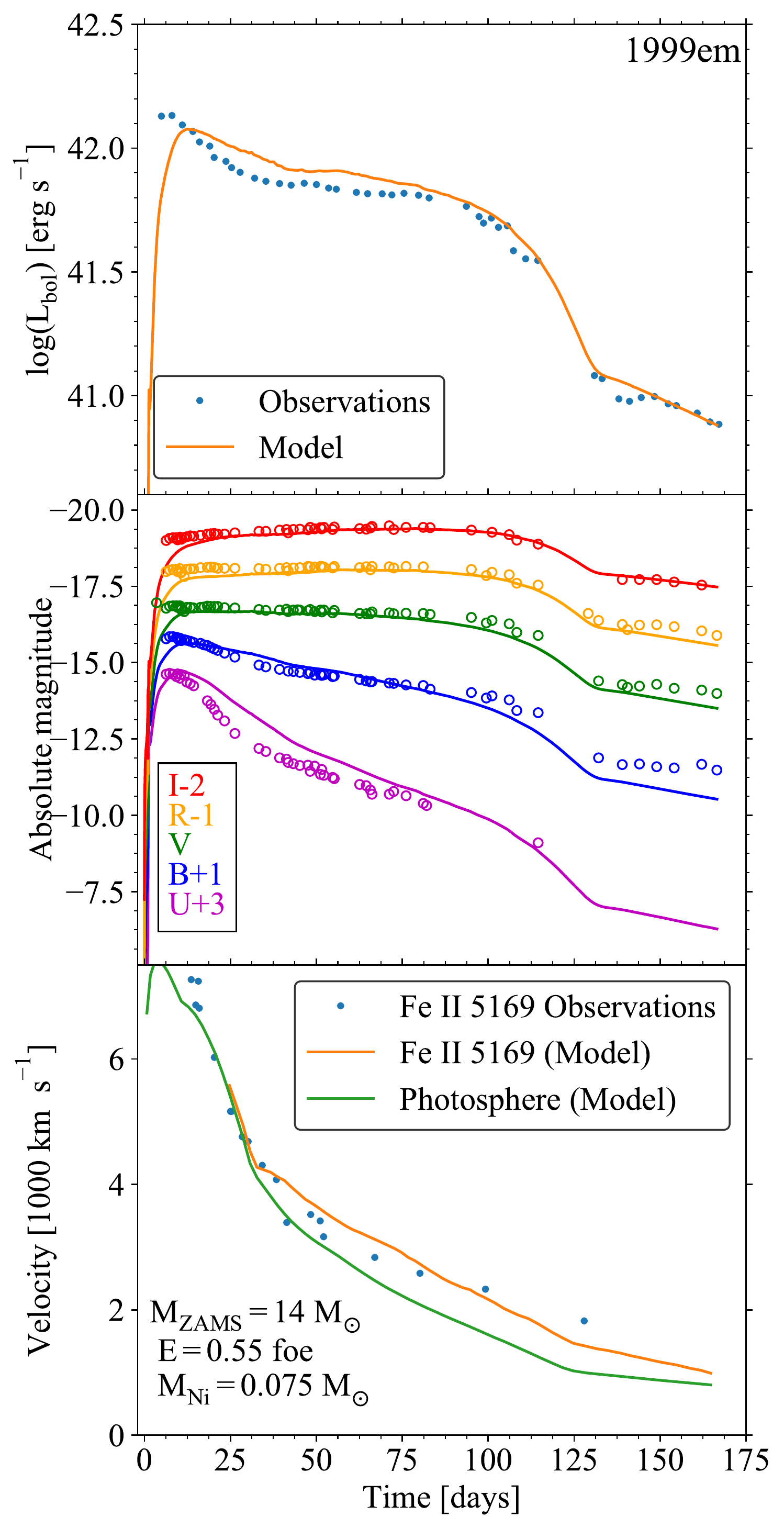}
\caption{Comparison of the best-fit MESA model with observational data
  for SN 1999em. From top-to-bottom, the figure shows the time
  evolution of the UBVRI quasi-bolometric lightcurve, the individual
  U, B, V, R and I color curves, the Fe II 5169 velocity, and the
  photospheric velocities. { M$_{Ni}$ refers to the $^{56}$Ni
    mass.}
\label{fig:1999em}}
\end{figure}

SN 1999em is a typical Type IIP SN that was discovered on 1999 Oct
29.44 UT in NGC 1637. We use 1999 Oct 24 (JD 2451476.0) as the
explosion date, following \citet{elmhamdi03}. Its plateau luminosity
(Figure \ref{fig:1999em}) is comparable to the average Type IIP
(Figure \ref{fig:lcs}), and the plateau duration is typical of the SNe
presented here. We adopt a distance of 11.7 Mpc, and reddening
E$_{B-V} = 0.10$ mag, as found by \citet{leonard03} using the
standard-candle method. We assume the ratio of total-to-selected
extinction to be R$_V = 3.1$ \citep{Cardelli89}, giving a total
line-of-sight extinction of A$_V= 0.31$. We compare our model to UBVRI
observational data from \citet{leonard02} and \citet{elmhamdi03}, with
the quasi-bolometric lightcurve and Fe II 5169 velocity data
digitized from figures in \citet[][herafter H16]{huang16}. 

The best-fit {\sc MESA} model for this SN, shown in Figure
\ref{fig:1999em}, provides very good agreement with the observed
plateau and tail luminosities, as well as the plateau duration. We
find a ZAMS mass of 14 $\msun$ for the progenitor. Our results for the
progenitor mass are comparable with those obtained via optical
progenitor detection \citep{smartt09}, and those derived from X-ray
and radio lightcurves \citep{pooleyetal02}. The results are also
consistent with the mass (12-14 $\msun$) and explosion energy (0.5-1
$\times 10^{51}$ erg) calculated by \citet{elmhamdi03} using the
plateau brightness and duration and the expansion
velocity. \citet{elmhamdi03} however used a distance of 7.8 Mpc
calculated by the Expanding Photosphere Method. Their values for {
  $^{56}$Ni} mass (0.02 M$_\odot$) and presupernova radius (120-150
R$_\odot$) differ significantly from our results.

M18 found a ZAMS mass ranging from 20 to 21.5 $\msun$, and an
explosion energy of 0.47 $\pm$ 0.05 foe, depending on the amount of
$^{56}Ni$ mixed in. In this case our explosion energy is in agreement
with M18, but their ZAMS mass is 50\% higher. The reason for this
discrepancy between their modelling and ours is unclear, although it
is possible that our modelling of the photospheric velocity
contributes to our lower mass. It is clear from the results of
\citet{morozovaetal16} that there exist substantial difference between
the {\sc MESA} + {\sc SNEC} models as compared to the {\sc KEPLER} +
{\sc SNEC} models, and we expect that these differences are further
exaggerated when using {\sc MESA} + {\sc STELLA}. We remark on
  this further in \S 4.

\subsection{SN 2004et}
\begin{figure}[!htb]
\includegraphics[width=\columnwidth]{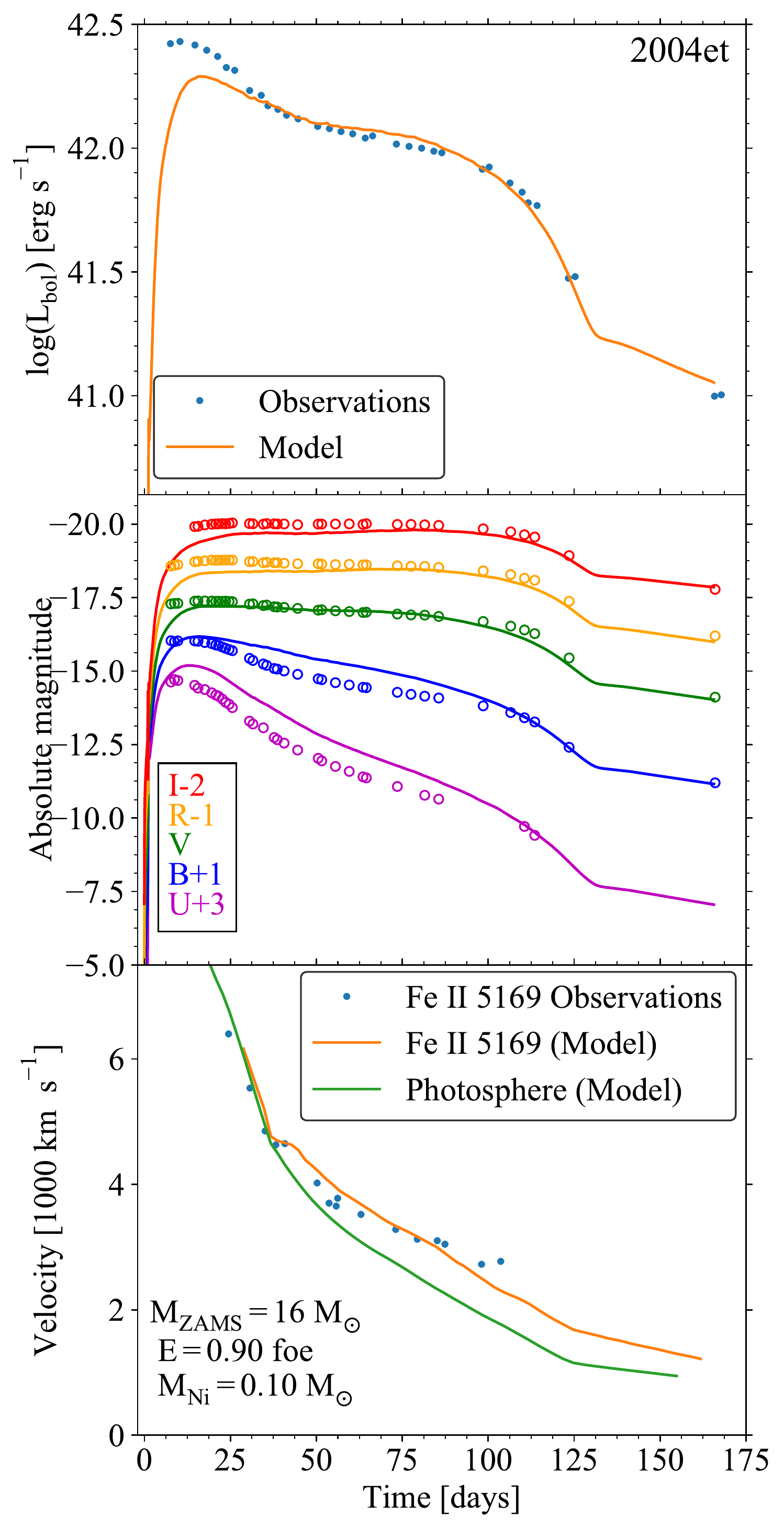}
\caption{Same as Figure \ref{fig:1999em}, but for SN 2004et
\label{fig:2004et}}
\end{figure}

SN 2004et was discovered by S. Moretti at about 12.8 mag in NGC 6946
on 2004 September 27. It is a bright SN, and one of the better studied
SNe included in this paper. We use 2004 September 22.0 (JD 2453270.5)
as the time of explosion for SN 2004et throughout this paper,
following \citet{lietal05}. They also found a distance of 5.5 Mpc and
extinction A$_v$ = 1.27 mag, which we adopt here. For this SN we use
observational data taken from \citet{sahu06}. Progenitor mass
estimates in the literature for this SN have encompassed a wide
range. Various estimates that have been mentioned include:
$15^{+5}_{-2}$ $\msun$ by \citet{lietal05} obtained by comparing the
intrinsic color and absolute magnitude to stellar evolutionary tracks;
10-20 $\msun$, but closer to 20 $\msun$, by \citet{misra07} using the
relations between various explosion parameters derived by \citet{ln85}
and \citet{popov93}; $\sim$ 20 $\msun$ by \citet{chevalier06}, by
modelling the radio and X-ray lightcurves; $27\pm2$ $\msun$ by
\citet{utrobin09} using hydrodynamic modelling; $9^{+5}_{-1}$ $\msun$
by \citet{smartt09} using direct optical progenitor detection; $<15$
$\msun$ by \citet{jerkstrand12} by modelling the late-time spectra;
and 16.5 $\msun$ by M18. Our best-fit model (Figure \ref{fig:2004et})
has a ZAMS mass of 16 $\msun$, falling in the middle of the deduced
range of values. It is in agreement with values obtained by
\citet{lietal05,misra07, jerkstrand12}, and M18, and close to the
value obtained from direct progenitor detection.  It is however lower
than that found by \citet{utrobin09}, who employed hydrodynamical
modelling. A possible reason is that \citet{utrobin09} used what they
term a non-evolutionary stellar model.  Our explosion energy and {
  $^{56}$Ni} mass estimates exceed those found by other
methods. However the good agreement between the simulated and observed
lightcurves using these parameters, at all but the earliest epochs,
suggests that our higher estimates are justified.
 
\subsection{SN 2009bw}
\begin{figure}[!htb]
\includegraphics[width=\columnwidth]{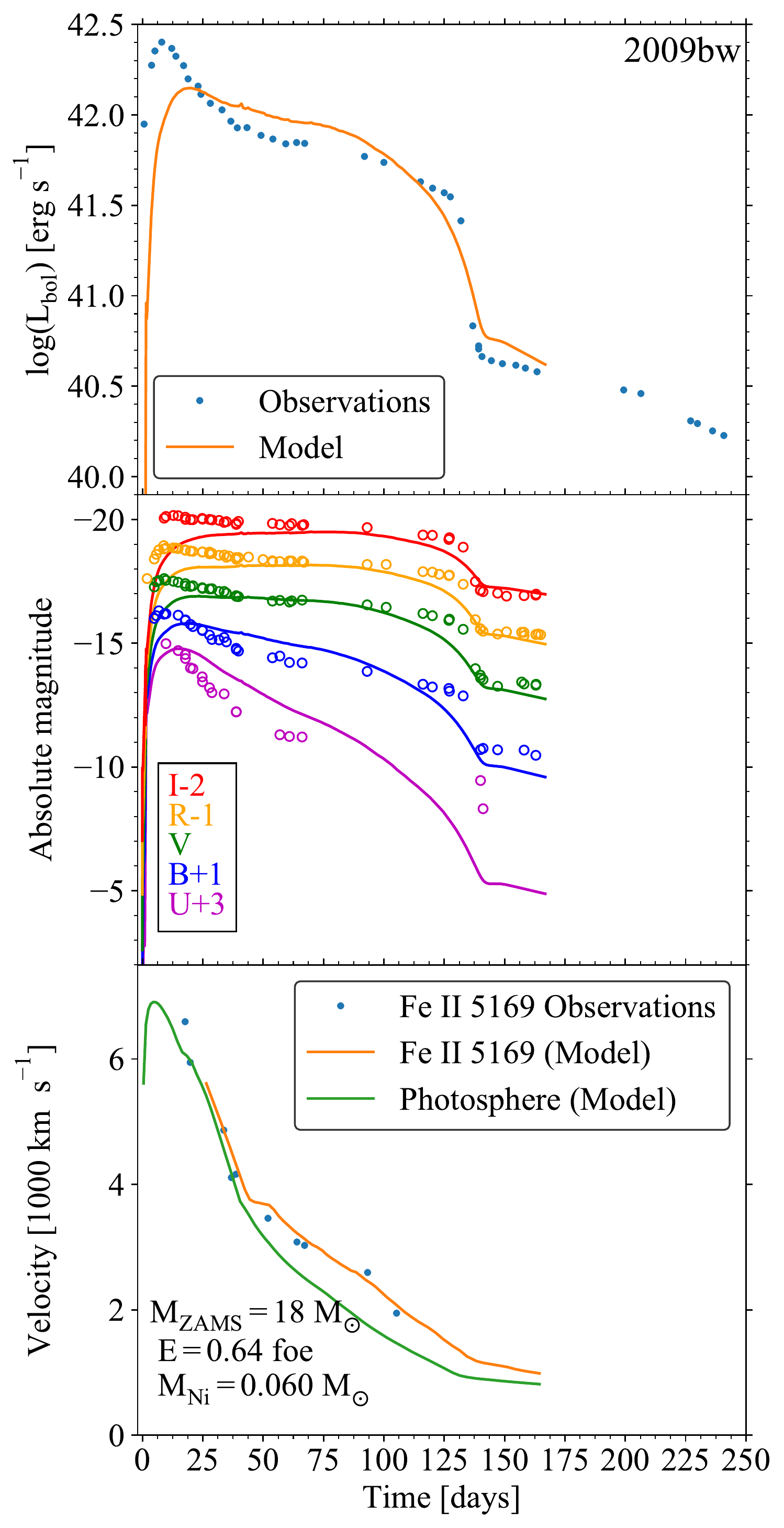}
\caption{Same as Figure \ref{fig:1999em}, but for SN 2009bw.
\label{fig:2009bw}}
\end{figure}

SN 2009bw has an unusual lightcurve, with a bright initial peak
followed by a long, flat plateau which falls off sharply at around 138
days post explosion. It was discovered in UGC 2890 on 2009 March
27.87, leading to an assumed explosion date of $\sim$ 2009 March 25
(JD 2454917.0) \citep{inserra12}. \citet{inserra12} used
hydrodynamical modeling to estimate an ejecta mass of 8.3-12 $\msun$,
an explosion energy of 0.3 foe, and a { $^{56}$Ni} mass of 0.022
$\msun$. We compare our model to observational data from this
paper. Our best-fit model (Figure \ref{fig:2009bw}), which adopts the
distance of 20.2 Mpc and total reddening E$_{B-V} = 0.31$ mag used by
\citet{inserra12}, gives an ejecta mass of 14.26 $\msun$ (from a ZAMS
mass of 18 $\msun$), an explosion energy of 0.64 foe and a {
  $^{56}$Ni} mass of 0.060 $\msun$. The fit to the quasi-bolometric
lightcurve is poor compared to some of the other SNe we investigate in
this paper, though the individual color curves and the Fe II
velocities also show a fairly good fit. We were unable to simulate the
early peak in the lightcurve alongside the less luminous plateau, even
with the addition of a large amount of CSM. The quick rise time is
also poorly reproduced by the model. The model does manage to
reproduce the plateau luminosity (especially in the V, R, and I
curves) and length, as well as the decline rate in the nebular phase
(specifically the late-time rate) and the Fe II velocities. We note
that the best fit obtained by \citet{inserra12} was equally poor, if
not worse. They have suggested that weak circumstellar interaction may
be playing a role in defining the lightcurve.

\subsection{SN 2013ab}
\begin{figure}[!htb]
\includegraphics[width=\columnwidth]{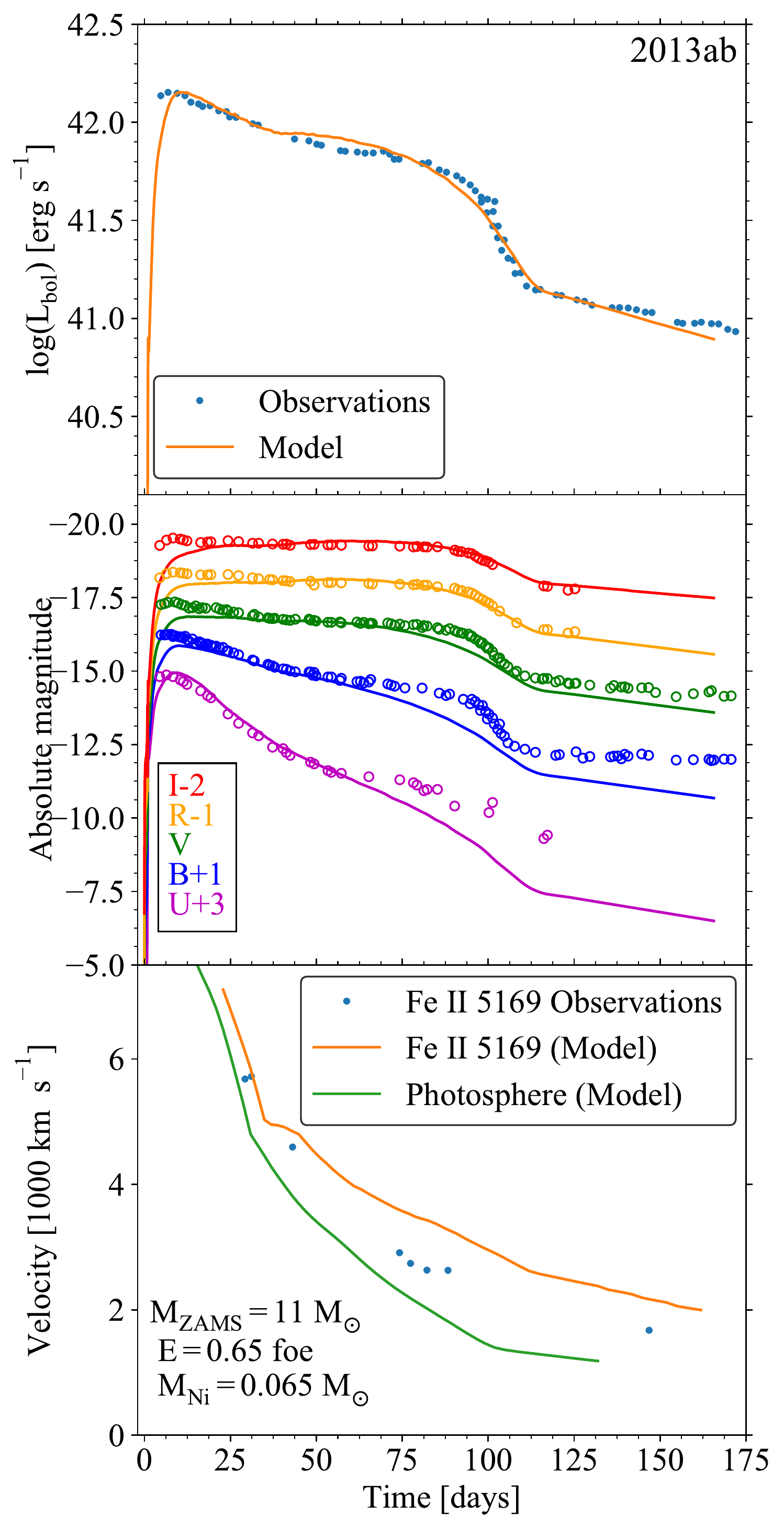}
\caption{Same as Figure \ref{fig:1999em}, but for SN 2013ab.
\label{fig:2013ab}}
\end{figure}

SN 2013ab was discovered on 2013 February 17.5, by
\citet{blanchardetal13} in the galaxy NGC 5669. An explosion date of
2013 February 16.5 (JD 2456340.0) is adopted from \citet{bose15}, who
assume a distance to the galaxy of 24 Mpc, with a total extinction
A$_v$ = 0.14 mag. All observational data are taken from
\citet{bose15}. The SN exhibits a noticeably similar lightcurve to
that of SN 1999em, with a slightly shorter plateau phase. Its Fe II
5169\AA\, velocity is also somewhat higher than that of SN 1999em for
the first 80 days. \citet{bose15} calculated a total { $^{56}$Ni}
mass of 0.064 $\msun$ by comparison to SN 1987A, and from the tail
luminosity \citep{hamuy03}. They then used a general relativistic,
radiation-hydrodynamical model to estimate the progenitor mass at
explosion of 9 $\msun$ and a radius of 600 R$_\odot$, with an
explosion energy of 0.35 foe. M18 found a progenitor ZAMS mass of 11.5
$\msun$ and an explosion energy of 0.84 foe, although they did allow
for a somewhat lower energy with a different degree of { $^{56}$Ni}
mixing. Our values are in good agreement with these, though the
explosion energy in both our (0.65 foe) and M18 calculations exceeds
that of \citet{bose15}. Our energy value arises primarily from fitting
the Fe II velocity. \citet{bose15} used the Sc II lines to obtain the
photospheric velocity, which is reasonable, but their best fit model
(Figure 18 in their paper) clearly underestimates the velocity. They
do consider a higher energy (0.6 foe, which would agree more with
ours) to better match the velocities, but find that it makes the light
curve fit much worse by lengthening the plateau phase. In our
simulated model the progenitor star is rotating at 35\% of the
critical velocity, an assumption that was not made in other
analyses. The resulting model fits the observations remarkably well
(Figure \ref{fig:2013ab}).

\subsection{SN 2014cx}
\begin{figure}[!htb]
\includegraphics[width=\columnwidth]{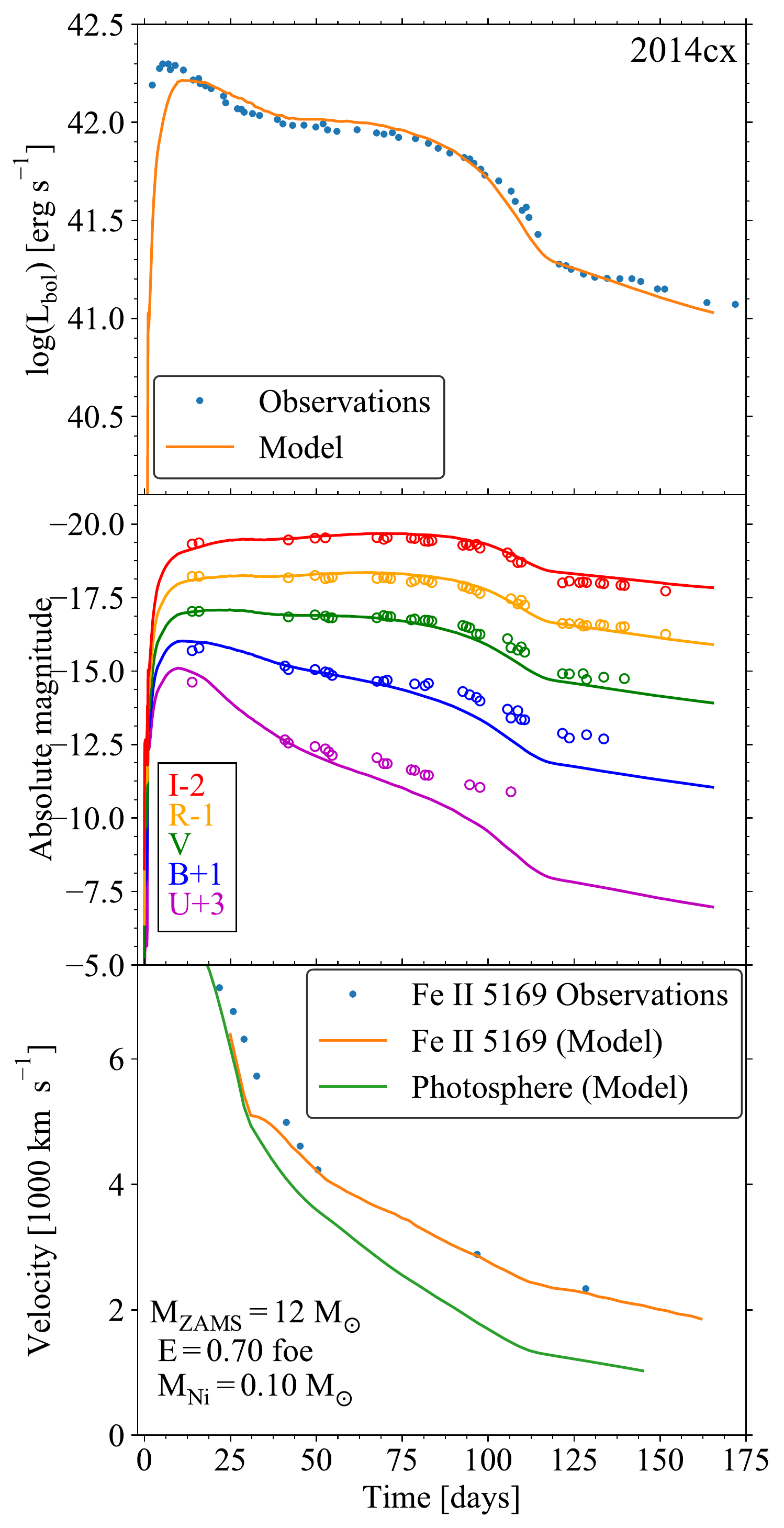}
\caption{Same as Figure \ref{fig:1999em}, but for SN 2014cx.
\label{fig:2014cx}}
\end{figure}

SN 2014cx (ASASSN-14gm) was discovered on UT 2014 September 2 in NGC
337 by \citet{holoienetal14} and \citet{nakanoetal14}.  It was likely
discovered within 1 day after the explosion, and has an estimated
explosion time of JD 2456902.4. Classification as a Type IIP followed
by \citet{eliasrosaetal14} and \citet{andrewsetal15}.  We adopt for
the host galaxy a distance of 18 $\pm$ 3.6 Mpc, and a total extinction
of A$_v$ = 0.31 mag, as used in H16. H16 deduced a { $^{56}$Ni}
mass of 0.056 $\pm$ 0.008 $\msun$ from comparison to SN 1987A, or
using the analytical formula derived by \citet{hamuy03}. Using
hydrodynamical modelling, they found a mass at explosion of $\sim$ 10
$\msun$ and a radius of 680 R$_\odot$, with an explosion energy of 0.4
foe. On the other hand, also using hydrodynamic modelling, M18 found a
ZAMS mass of $>$ 22 $\msun$ and an energy of 0.66 $\pm$ 0.04 foe. Our
simulated model, compared to the observational data from H16,
reproduces the lower mass estimate of H16. However we find that a
higher energy, as given by M18, is necessary to match the Fe
velocities. We note that the model fit of H16 (Figure 12 in their
paper) underestimates the photospheric velocity, especially in the
first two months. Their model fit to the bolometric lightcurve is also
not convincing. Our model (Figure \ref{fig:2014cx}) requires an
especially high { $^{56}$Ni} mass to fit the nebular phase of the
lightcurves. It fits the data well, with the exception of a slow rise
time. This lightcurve is notable for having a short plateau, which in
our stellar evolution model could only be adequately fit with a
progenitor rotating at about a third of the critical velocity.

\subsection{ASASSN-14dq}
\begin{figure}[!htb]
\includegraphics[width=\columnwidth]{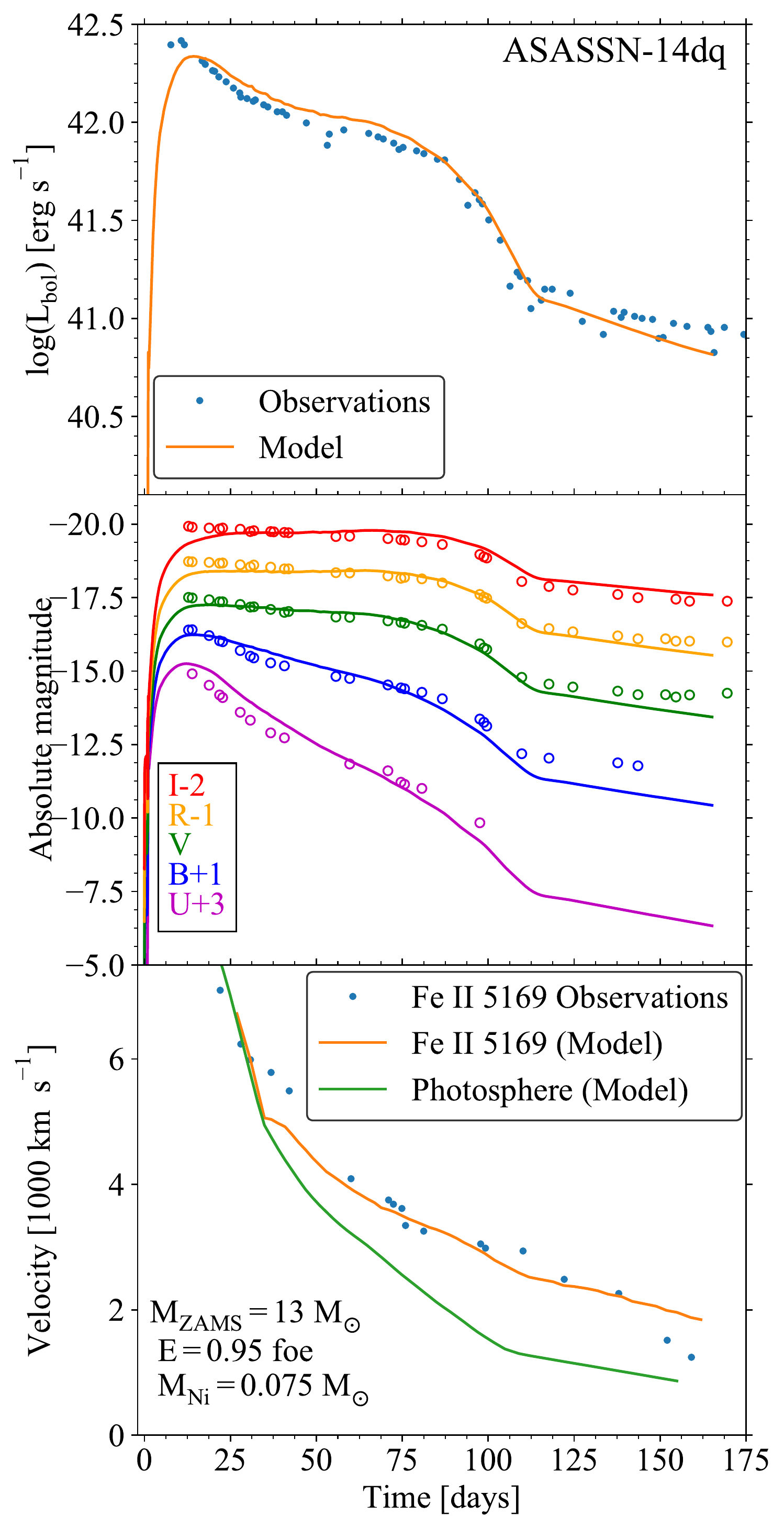}
\caption{Same as Figure \ref{fig:1999em}, but for ASASSN-14dq
\label{fig:14dq}}
\end{figure}

ASASSN-14dq was discovered on 2014 UT July 08.48 in the low-luminosity
dwarf galaxy UGC 11860 \citep{staneketal14}. Observational data are
taken from \citet {singh18}. \citet{singh18} estimated a mean distance
to the host galaxy of 44.8 Mpc using the standard candle method, with
a total reddening E$_{B-V} = 0.06$ mag. Given the lack of existing
measurements of the metallicity of this galaxy, \citet{singh18}
estimated the metallicity using various luminosity-metallicity
relations. They found a sub-solar oxygen abundance for the host
galaxy. Using model spectra generated for four 15 M$_\odot$ SN
progenitors \citep{dessartetal13} with metallicities of 0.1, 0.4, 1,
and 2 Z$_\odot$, \citet{singh18} showed that the spectra matched
closely with those at a metallicity Z $=0.4$ Z$_\odot$.  Using an
analytic light-curve model, they estimated the ejecta mass from this
SN to be $\approx$ 10 $\msun$, with an explosion energy of 1.8 foe,
and a total { $^{56}$Ni} mass of 0.029 $\msun$. M18 found a
progenitor ZAMS mass of 18.5-19.5 $\msun$ and an energy of 0.86
foe. We were unable to reproduce the lightcurves using a stellar
evolution model with a metallicity of 0.4 Z$_\odot$. Our best-fit
model (Figure \ref{fig:14dq}) instead has Z $=0.7$ Z$_\odot$.  Given
the gap between the 0.4 Z$_\odot$ and 1 Z$_\odot$ data points in the
model spectra used by \citet{singh18}, and the difference in mass
between our progenitor and the 15 M$_\odot$ model, a metallicity of
0.7 Z$_\odot$ is quite plausible. Our model has a low ejecta mass,
comparable to that found by \citet{singh18}. However our model does
not need an exceptionally high explosion energy, with the best fit
giving 0.95 foe, about half the explosion energy suggested by
\citet{singh18}, but consistent with that of M18. It does need a
$^{56}$Ni mass about 2.5 times higher than that found by
\citet{singh18}. It is however to be noted that \citet{singh18} have
found that their derived $^{56}$Ni mass was much lower than expected
for the plateau duration of the SN.

\subsection{SN 2015ba}
\begin{figure}[!htb]
\includegraphics[width=\columnwidth]{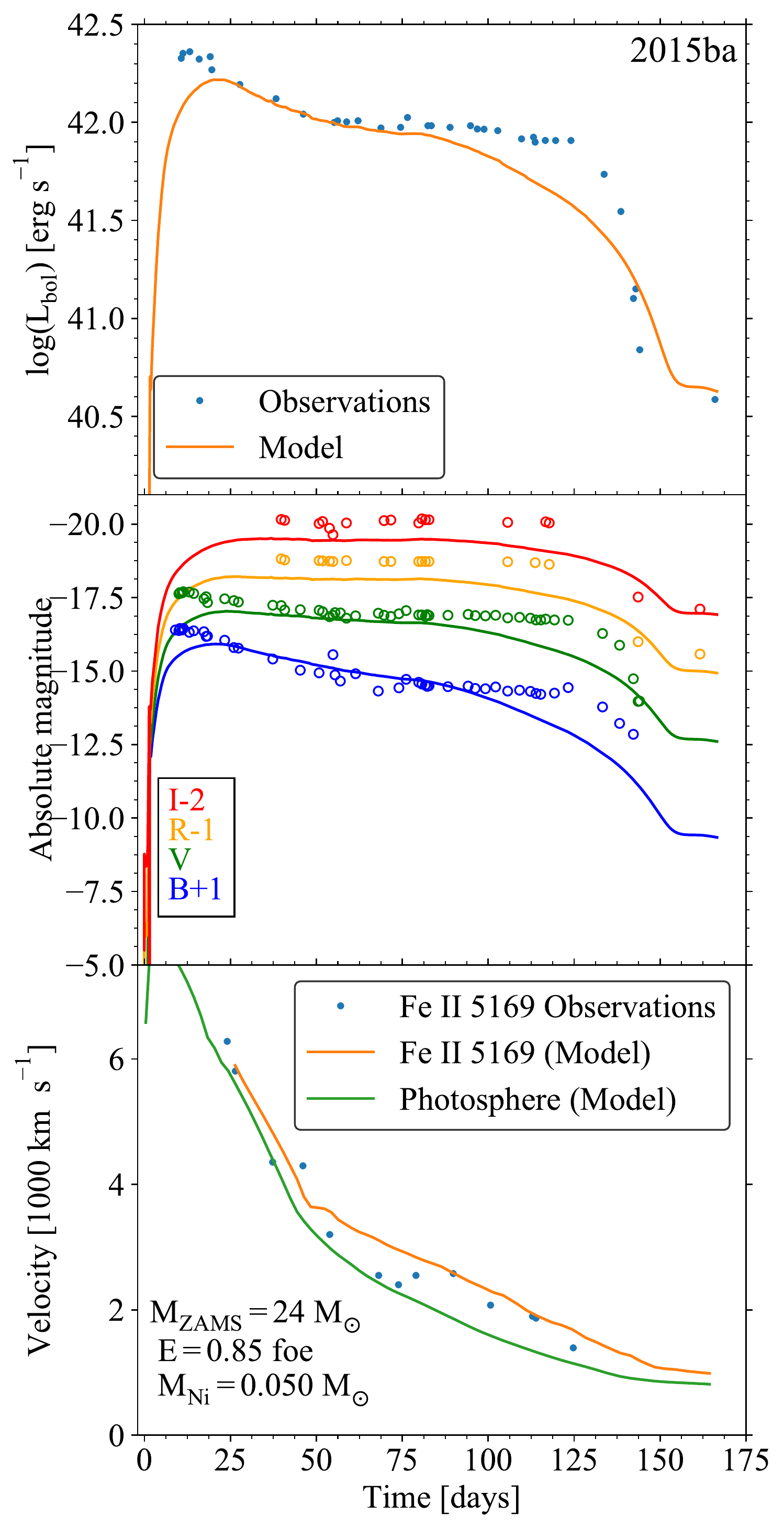}
\caption{Same as Figure \ref{fig:1999em}, but for SN 2015ba. In this case only B, V, R
  and I color curves are available, and comparing the UBVRI
  quasi-bolometric curve from MESA to the BV\textit{ri}
  quasi-bolometric curve derived from observations.
\label{fig:2015ba}}
\end{figure}

SN 2015ba was discovered on 2015 November 28.8071 UT in the galaxy IC
1029. \citet{dastidar18} used a cross-correlation technique to
determine the epoch of explosion, finally settling on a date 2015
November 23 (JD 2457349.7 $\pm$ 1.0) as the explosion date. Using a
weighted mean of distances, they adopt a distance of 34.8 $\pm$ 0.7
Mpc, and use a total reddening value of E$_{B-V}$= 0.46 mag.  We use
their values in this paper. The observational data referenced here
also comes from \citet{dastidar18}, but is less complete than that of
the other SNe studied in this paper. U-band observations are missing,
and RI data are incomplete, so the quasi-bolometric lightcurve is
calculated using the BV\textit{ri} bands rather than the standard
UBVRI. This SN is notable for an unusually long, flat plateau, which
ends in a sharp drop of $\sim$3 magnitude. In fact, there does not appear
to be another well-studied Type IIP SN with a plateau as luminous and
long as that of SN 2015ba \citep{anderson14, dastidar18}.

\citet{dastidar18} find a { $^{56}$Ni} mass for this SN of 0.032
$\msun$. Using analytical and general-relativistic, radiation
hydrodynamical modeling, they estimate the ejecta mass at 22-24
$\msun$ and the explosion energy at 1.8-2.3 foe. In our modeling we
find that a similarly large progenitor mass is necessary in order to
produce a long lightcurve with a large drop in luminosity in the
nebular phase. However, even using larger progenitor masses, we were
unable to fit the observed lightcurves for SN 2015ba with the same
precision that we obtained for the other SNe in this paper.  Our best
model, with a mass of 24 $\msun$, an explosion energy of 0.85 foe, and
a { $^{56}$Ni} mass of 0.050 $\msun$, still does a poor job of
modeling the sharp drop-off from the plateau to the nebular phase (see
Figure \ref{fig:2015ba}). A small part of this inconsistency may be
due to our comparison of the UBVRI quasi-bolometric lightcurve
generated by {\sc STELLA} to the BV\textit{ri} quasi-bolometric curve
derived from observations, but we do not expect this to be the major
issue. U-band lightcurves tend to fall off more gradually than the
other bands, due to cooling of the material and degradation of
photons. Given this shortcoming, our best-fit parameters for SN 2015ba
have greater uncertainty than those for the other SNe modeled in this
paper. What is clear though is that the progenitor mass is very high
compared to those of the other SNe, and likely in the same ballpark as
that estimated by \citet{dastidar18}.

\subsection{SN 2016X}
\begin{figure}[!htb]
\includegraphics[width=\columnwidth]{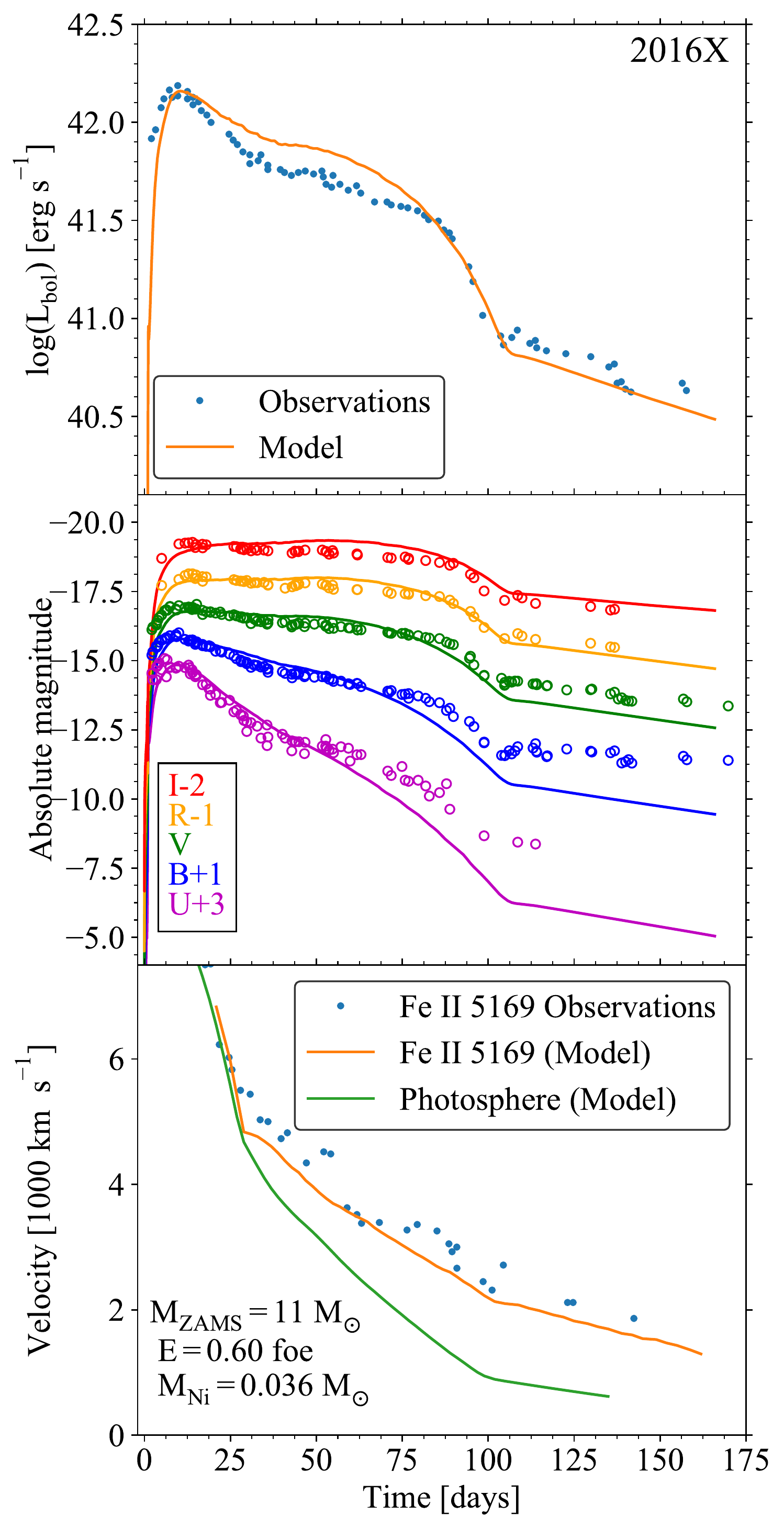}
\caption{Same as Figure \ref{fig:1999em}, but for SN 2016X
\label{fig:2016X}}
\end{figure}

Of all the SNe investigated in this paper, SN 2016X is the least
luminous, and has the shortest plateau. It was discovered by the All
Sky Automated Survey for SuperNovae (ASAS-SN) on 2016 January 20.59 UT
in the nearby SBd galaxy UGC 08041 (z = 0.004 408 from NED). An
explosion date of 2016 January 18.9 (JD 2457406.4) was adopted by
\citet{huang18}. All observational data referenced here is taken from
\citet{huang18}. They find a distance to the host galaxy of 15.2 Mpc
using the Tully-Fisher method, and a total reddening of E$_{B-V}$ =
0.04 mag. Based on the photospheric temperature, they estimate the
radius of the immediate progenitor at 860-990 R$_\odot$, corresponding
to a mass of 18.5-19.7 $\msun$. They find a { $^{56}$Ni} mass of
0.034 $\msun$ by comparison to SN 1987A. Our best fit to the
lightcurves (Figure \ref{fig:2016X}) gives significantly different
parameters, with a ZAMS mass of only 11 $\msun$ and a radius of 536
R$_\odot$. Our value for { $^{56}$Ni} mass agrees well with the
estimate of \citet{huang18}. We find that the quasi-bolometric fit for
this SN is poorer than that of many others examined in this paper. The
plateau luminosity would appear to suggest a lower explosion energy,
but the relatively large Fe velocities suggest the opposite. The
comparison shown in Figure \ref{fig:2016X} represents a compromise
between these two fitting parameters. Despite the difficulty in
matching the quasi-bolometric lightcurve, the individual UBVRI
lightcurves compare fairly well, and clearly require a progenitor mass
well below the estimate of \citet{huang18}.

\ \\
\section{Summary and Discussion}
\subsection{Quality of the Fits}
\begin{figure}[!htb]
\includegraphics[width=\columnwidth]{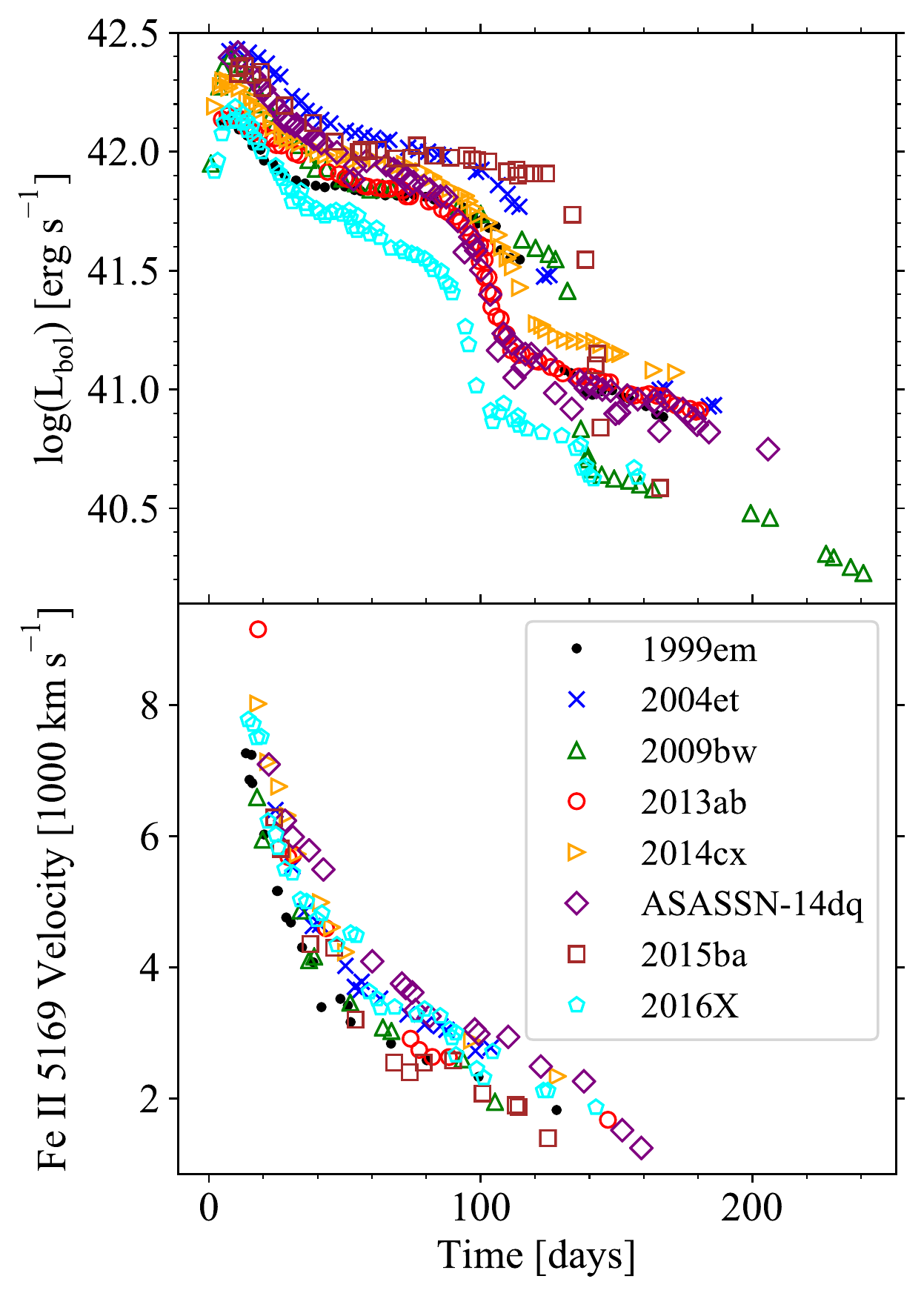}
\caption{Comparison of (top) the quasi-bolometric lightcurves and
  (bottom) the Fe II 5169 velocities for SN 1999em \citep{leonard02,
    elmhamdi03}, SN 2004et \citep{sahu06}, SN 2009bw
  \citep{inserra12}, SN 2013ab \citep{bose15}, SN 2014cx
  (H16), ASASSN-14dq \citep{singh18}, SN 2015ba
  \citep{dastidar18}, and SN 2016X \citep{huang18}.\label{fig:general}
\label{fig:lcs}}
\end{figure}

In this paper we have simulated the quasi-bolometric lightcurves, the
UBVRI color lightcurves and the photospheric velocities for eight SNe,
and compared these to the observational data. We have used the
best-fit models to determine the properties of the explosion, such as
the explosion energy and { $^{56}$Ni} mass, as well as the
properties of the progenitor star, in particular the ZAMS progenitor
mass.  The quasi-bolometric lightcurves and Fe II 5169\AA\,velocities
for all the SNe covered in this paper are shown in Figure
\ref{fig:lcs}.

Overall, we find that a good fit to the quasi-bolometric curve
typically results in a good fit to the color curves, though this is
less true in the cases of SNe 2009bw and 2016X. The U and B fits tend
to be poorer than the V, R and I, though these are still fairly good
in most cases.

\begin{figure}[!htb]
\includegraphics[width=\columnwidth]{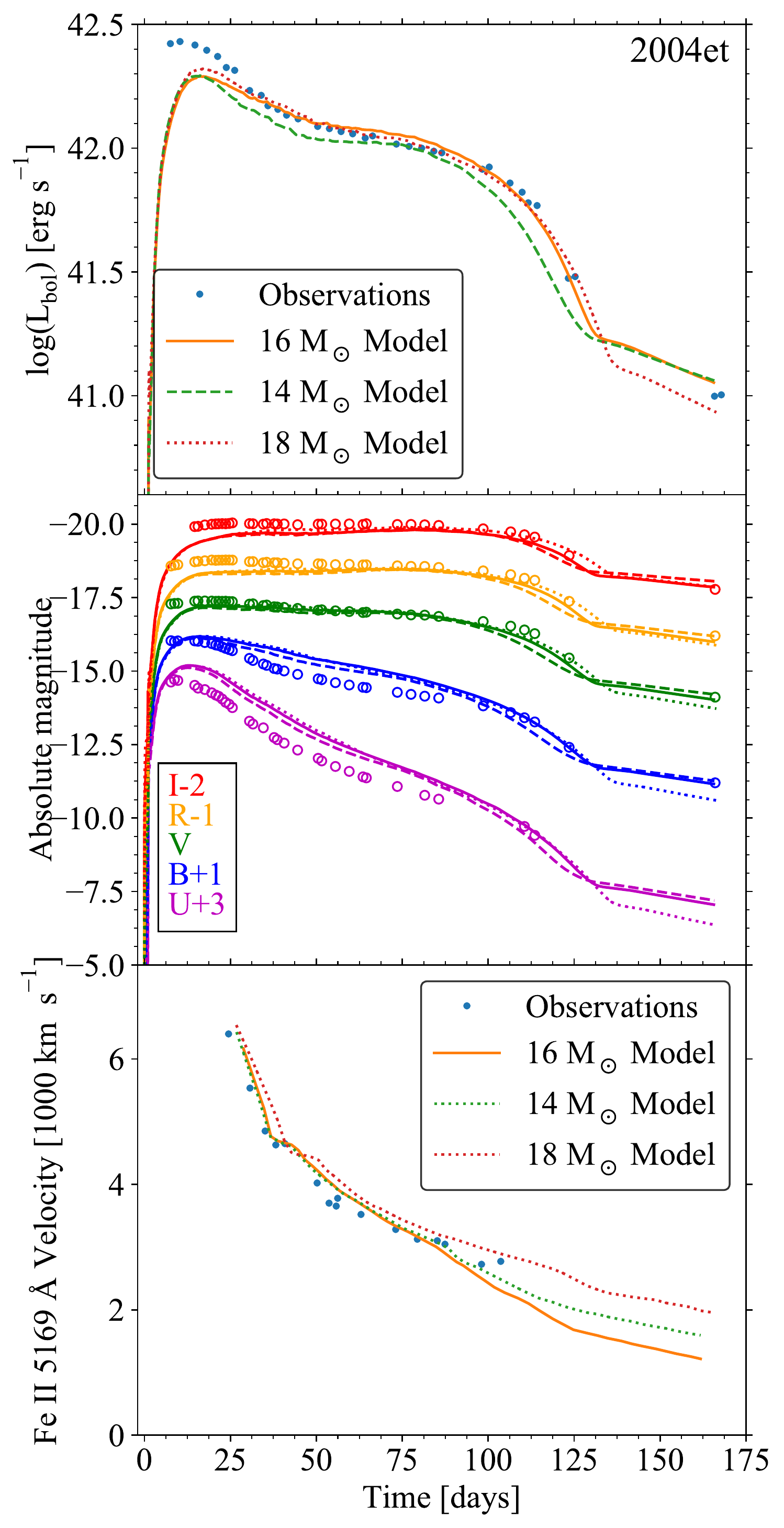}
\caption{Same as Figure \ref{fig:1999em}, but for SN 2004et. Models
  identical to the best-fit in all parameters except ZAMS mass are
  shown as dashed (14 $\msun$) and dotted (18 $\msun$) lines.
  Photospheric velocities are not shown.\label{fig:general}
\label{fig:alt}}
\end{figure}

There is undoubtedly some leeway in the parameters for each SN,
although the specific error in each case is difficult to
quantify. Given the good fit in most cases, we suspect that this
uncertainty is quite low, and we are confident that in most cases the
progenitor masses are determined to within about a solar mass. Figure
\ref{fig:alt} gives the best-fit model for SN 2004et alongside
otherwise identical models with $\pm$2 $\msun$ at ZAMS. These models
have clearly diverged from the best-fit: the 14 $\msun$ model has good
Fe II 5169 velocity agreement but produces a less luminous and shorter
plateau. Adjusting other parameters does not improve the fit.  The
larger 18 $\msun$ model does appear to fit the lightcurve reasonably,
but gives Fe II 5169 velocities that are too high. Adjusting these to
be more in line with the data would require lowering the explosion
energy, which would in turn increase plateau length, thereby
destroying the lightcurve fit.

There do not appear to be any significant degeneracies, i.e. models
with similar features but drastically different parameters, among our
model fits. Requiring matches to all of the bolometric lightcurve,
UBVRI color curves and Fe II velocities helps to eliminate some
degeneracy, as exemplified by the models mentioned above.
\citet{goldberg19} find that some degeneracy exists between models
with various initial masses in {\sc MESA/STELLA} when matching both
the bolometric lightcurves and the Fe II velocities. However they did
not additionally compare the UBVRI color lightcurves, which would
certainly help.  Furthermore, they argue that early-time Fe
velocities, which are shown to vary greatly based on explosion energy
and the compactness of the progenitor star, can be used to eliminate
the degeneracy in cases where there is not substantial CSM
present. Given the difficulty in obtaining the early Fe II velocities
mentioned in section \ref{sec:mesa}, we have used the photospheric
velocity at early times instead. All of our models succeed in matching
the early SN velocities, and we note that these early velocities help
us to eliminate degenerate models even with some CSM present.

In their comparison of Type IIP SN lightcurves using the {\sc KEPLER}
and {\sc SNEC} codes, M18 often found much higher progenitor masses
than we find in this paper. However they did not compute the Fe II
5169 velocities, which, as shown in the above example, are essential
to breaking the degeneracy in progenitor mass and explosion energy.

The SN progenitor models produced using {\sc MESA} and {\sc STELLA}
were found to be quite capable of reproducing the lightcurves and
photospheric velocities of Type IIP SNe. We used the sample routines
as described in \citet{mesa4} to carry out the stellar evolution
modelling and the SN explosion, modifying a minimal number of
parameters required to get the evolution correct. We did not change
the technique employed to explode the SN. In our work, we found that
there are certain features of the lightcurves which {\sc STELLA} seems
to have difficulty with. The U and B color curves tend to fit much
worse than the V, R and I curves, especially at later times. (We
  address this in detail at the end of this section.) Furthermore,
the rise time in our models is generally too long. Some of this may be
due to the uncertainty in the explosion times. M18 make the
observation that {\sc STELLA} produces rise times 3-5 days longer than
those obtained using {\sc SNEC}, which we appear to see in our
models. Additionally early peaks in the lightcurves tend to be a
problem for {\sc STELLA}, especially when using a large number of
frequency bins. Although the addition of a large amount of CSM does
help in this regard, it often proves inadequate, such as for SNe
2004et, 2009bw and 2015ba.

We furthermore found the modelling of several sub-luminous SNe,
including SN 2009N and SN 2005cs, to be problematic using the
techniques outlined herein. Reproducing the dim plateaus of these SNe
proved difficult without the use of exceedingly low explosion
energies, which resulted in significantly lower Fe II 5169
velocities. Adjustments to mixing length and overshooting appear to be
promising avenues towards reproducing these sub-luminous lightcurves,
and will be explored in future work. Notwithstanding all this, {\sc
  STELLA} generally does an excellent job reproducing the most
important features of standard SN lightcurves, and is an excellent
addition to the suite of tools available to a SN astronomer.

The majority of the fits presented in this paper indicate SN
progenitor masses that overlap with some of the previous estimates,
with the exception of SN 2016X. We find that the {\sc MESA} and {\sc
  STELLA} combination used in this paper tends to give lower
progenitor masses than previous hydrodynamical modeling techniques
(see \S 3.2), and typically agrees with the general-relativistic,
radiation-hydrodynamical code that has been used to model SNe 2009bw,
2013ab, 2014cx and 2015ba by \citet{inserra12}, \citet{bose15},
H16, and \citet{dastidar18}, respectively. The modeling
done by M18 using {\sc SNEC} agrees with our own in the cases of SNe
2004et and 2013ab, but produces significantly higher progenitor masses
for SNe 1999em, 2014cx, and ASASSN-14dq. 

\begin{figure}[!htb]
\includegraphics[width=\columnwidth]{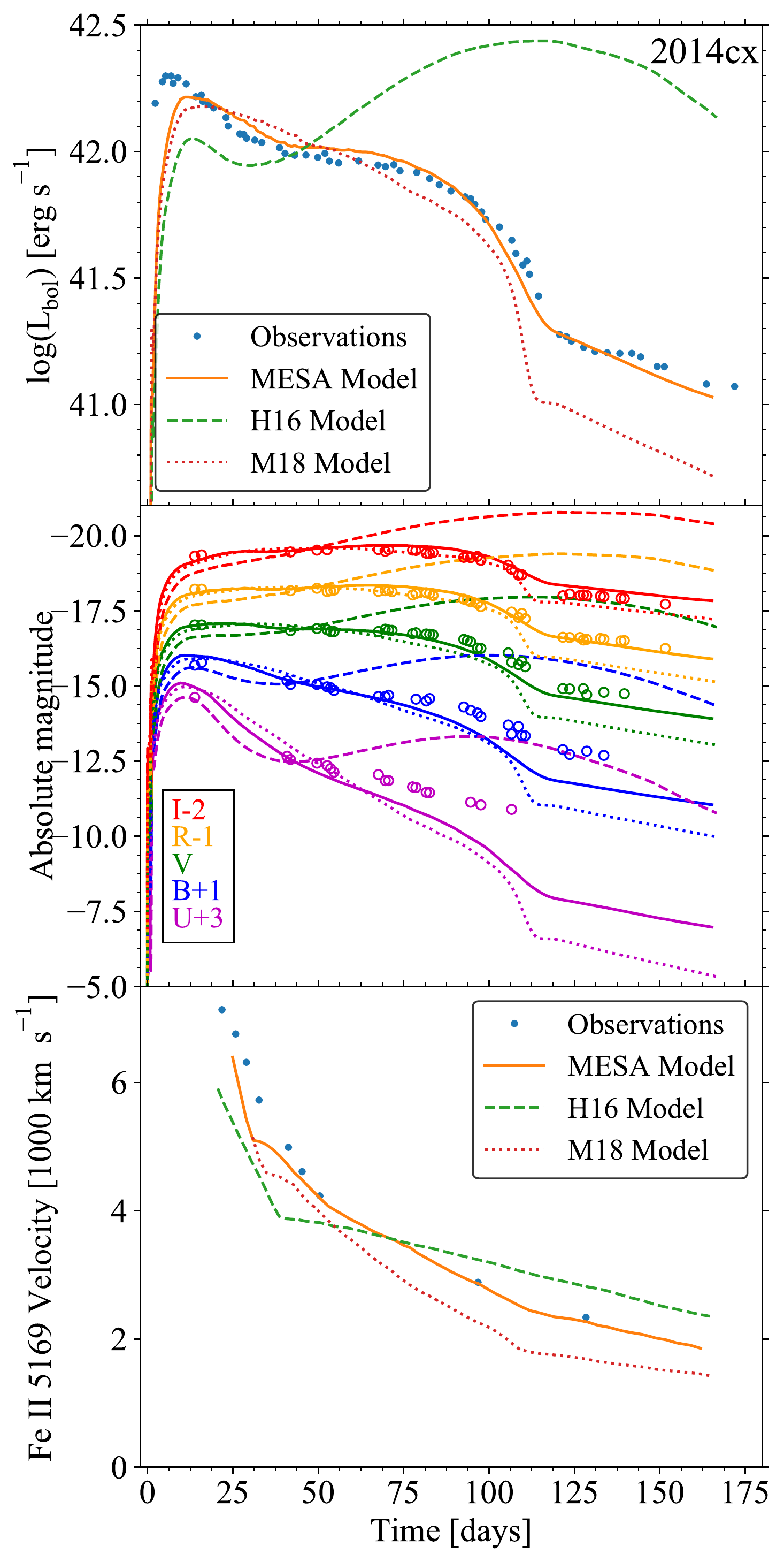}
\caption{Same as Figure \ref{fig:1999em}, but for SN 2014cx. Our
  best-fit model, as given in \S 3.5, is shown in solid lines. Models
  created with {\sc MESA} and {\sc STELLA} using the best-fit
  parameters of H16 and M18 are shown as dashed and dotted lines,
  respectively. Photospheric velocities are not shown.
\label{fig:mlt}}
\end{figure}

In order to study the differences between our model fits and those
calculated by others, we have input the best-fit parameters obtained
by other codes in our MESA/STELLA combination. Figure \ref{fig:mlt}
shows our best-fit model for SN 2014cx alongside models produced in
{\sc MESA} using the explosion parameters derived by H16 and M18, as
given in \S 3.5. While the original {\sc KEPLER/SNEC} model presented
in M18 lines up quite well with observational data, our M18-inspired
MESA model shows a much larger discrepancy with the observational
data. { $^{56}$Ni} mixing, discussed in detail below, may account
for some of this divergence. In the case of H16, the original
general-relativistic, radiation-hydrodynamics code model presented in
their paper did not fit the data as accurately as ours or
M18's. However, the {\sc MESA} recreation of this model does not even
properly reproduce the shape of the lightcurve. We were forced to
reuse our rotating 12\msun\ model in this case, as {\sc MESA} failed
to explode a similarly sized non-rotating model, but the presence of
rotation should not cause nearly the degree of variation we observe
here. Due to the complexity of these codes and the large number of
parameters involved, the reasons for such significant disagreement
between the three remain unclear. However, it is clear that the
differences in the best-fit parameters arise to a substantial degree
due to the differences in the codes themselves, both in the stellar
evolution modelling and the light-curve modelling. An unfortunate
conclusion from this may be that results from all codes are
suspect. However, in our opinion the results from {\sc MESA} + {\sc
  STELLA} appear to be more in line with observations. To be certain,
MESA also has its shortcomings, but the large user base makes it
likely that problems will be caught early. To the developers credit,
they are continually and actively working to add more physics modules,
while improving the code and fixing errors, as the series of MESA
papers shows.

\begin{figure*}[!htb]
\includegraphics[width=\textwidth]{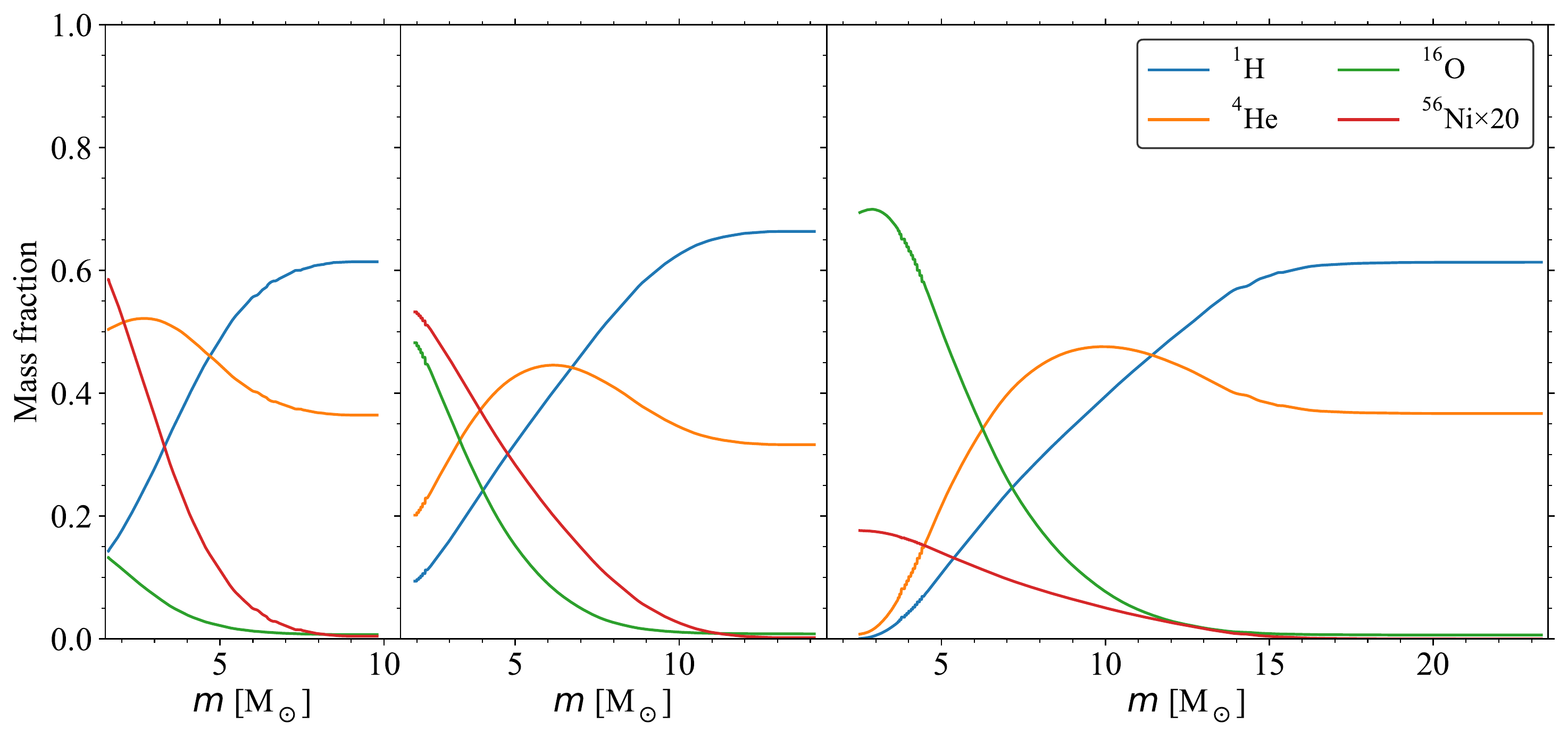}
\caption{Abundances of key species at 50 days post explosion
    for our best-fit models of SN 2013ab (11 $\msun$, left), SN 2014cx (16 $\msun$, center), and SN 2015ba (24 $\msun$, right).\label{fig:general}
\label{fig:abn}}
\end{figure*}

In the past, hydrodynamical models of Type IIP SNe generally required
much higher progenitor ZAMS masses. This has been attributed to the
use of non-evolutionary models, or spherically symmetric models that
do not take hydrodynamic instabilities and mixing into account. Both
these problems are rectified in this version of {\sc MESA} and {\sc
  STELLA} to some degree. Although the SN explosion models are
one-dimensional, they do take the multi-dimensional effects of the
Rayleigh-Taylor instability and mixing into account using the
prescription by \citet{duffell16}. The Duffell RTI scheme is applied
by {\sc MESA} to $^{56}$Ni mixing during the SN explosion. The degree
of { $^{56}$Ni} mixing has been shown to have a significant effect
on SN lightcurves, and this process has typically been handled in the
past by adding { $^{56}$Ni} uniformly out to a chosen mass
coordinate \citep{bersten11, morozova18}. {\sc MESA} takes a different
approach, first adding { $^{56}$Ni} uniformly out to a mass
coordinate consistent with 3D simulations (see Figure 27 in
\citet{mesa4}) just before the shock reaches the H shell. The {
  $^{56}$Ni} distribution is then allowed to evolve through the
Duffell RTI until just before shock breakout, at which point the
resulting distribution is rescaled in place to match the chosen total
{ $^{56}$Ni} mass. This dynamic process produces a smoothly mixed
final distributions like those shown in Figure \ref{fig:abn}, which
vary as expected with ZAMS mass but are generally consistent. In our
runs we have not altered the default parameters used by {\sc MESA} for
the Duffel RTI, since it was shown in \citet{mesa4} that the effect is
relatively small.

As noted above, the U and B color curves in our models often fall off
faster than the observations at later epochs. A similar behavior was
noted by \citet{blinnikovetal06} when using STELLA. They attributed it
to a large degree of { $^{56}$Ni} mixing, which leads to a more
rapid evolution. This is exactly the behavior that we see for instance
in SN 2016X (Figure \ref{fig:2016X}, where the early-time U-band flux
matches the peak well, but decreases faster than the observed flux. In
the case of ASASSN-14dq (Figure \ref{fig:14dq}) the U-band flux is
somewhat higher than the observed flux at early times, but lower than
the observed one at late times. It is possible that reducing the
mixing may be beneficial to the late-time lightcurves, but may affect
the early time flux adversely, as well as the other colors.  It may be
necessary to somehow reduce the mixing proportionately in the outer
layers, which is beyond the scope of this work.

\subsection{Explosion and Progenitor Properties}
In \S \ref{sec:intro} the RSG problem was brought up, asserting that
RSGs which exploded to form Type IIP SNe had ZAMS progenitor masses
$\le 17-19 \msun$. Though the X-ray limit is not well constrained,
what is important is that a mass limit exists, which is lower than the
maximum mass of a RSG, as deduced from stellar evolution theory. A
partial explanation may arise from theoretical modelling, with
\citet{sukhboldetal16} claiming that only about 10\% of SNe arise from
stars $> 20 \msun$, and some of these are Type Ib or Ic. They find
that there are only small `islands' of progenitor masses above 20
$\msun$, where stars undergo core-collapse to form a SN.

For the most part our results are consistent with the
  assertions of \citet{sukhboldetal16}. 7 of our 8 SNe indicate
  progenitor masses lower than 18 $\msun$.  The fit to SN 2015ba
  suggests a ZAMS mass in excess of this value, although the poor
  quality of the fit does not allow for any firm conclusions,
  especially given that the lightcurve does not resemble the other
  Type IIP SN lightcurves studied herein (see Figure
  \ref{fig:lcs}). Given our small sample size, and the fact that many
  of these SNe were thought to have much larger progenitor masses in
  the past, 1 out of 8 or 12.5\% having a progenitor mass above 20
  $\msun$ is in keeping with theoretical expectations.

The { $^{56}$Ni} masses presented in this paper are obtained
directly from the fits, without appealing to (semi-)analytic values or
comparing to another SN such as SN 1987A. They appear well
constrained, as the { $^{56}$Ni} mass almost exclusively determines the
luminosity of the radioactive tail of the plateau for a given ZAMS
mass. Many of the { $^{56}$Ni} masses referenced in the literature are
derived by comparison to the bolometric lightcurve of SN 1987A, as in
\citet{hamuy03}. Our values are generally larger than those previously
found, with the exception of SNe 2013ab and 2016X. However, they still
show a direct correlation between { $^{56}$Ni} mass and tail luminosity.

\citet{muller17} calculated a relation between { $^{56}$Ni} mass and plateau luminosity:

\begin{equation}
\log(\frac{{ M_{^{56}Ni}}}{M_\odot})=1.55^{+0.16}_{-0.14}\,\log(\frac{L_{pl}}{L_\odot})-14.51^{+1.31}_{-1.24}
\end{equation}

\begin{figure}[!htb] 
\includegraphics[width=\columnwidth]{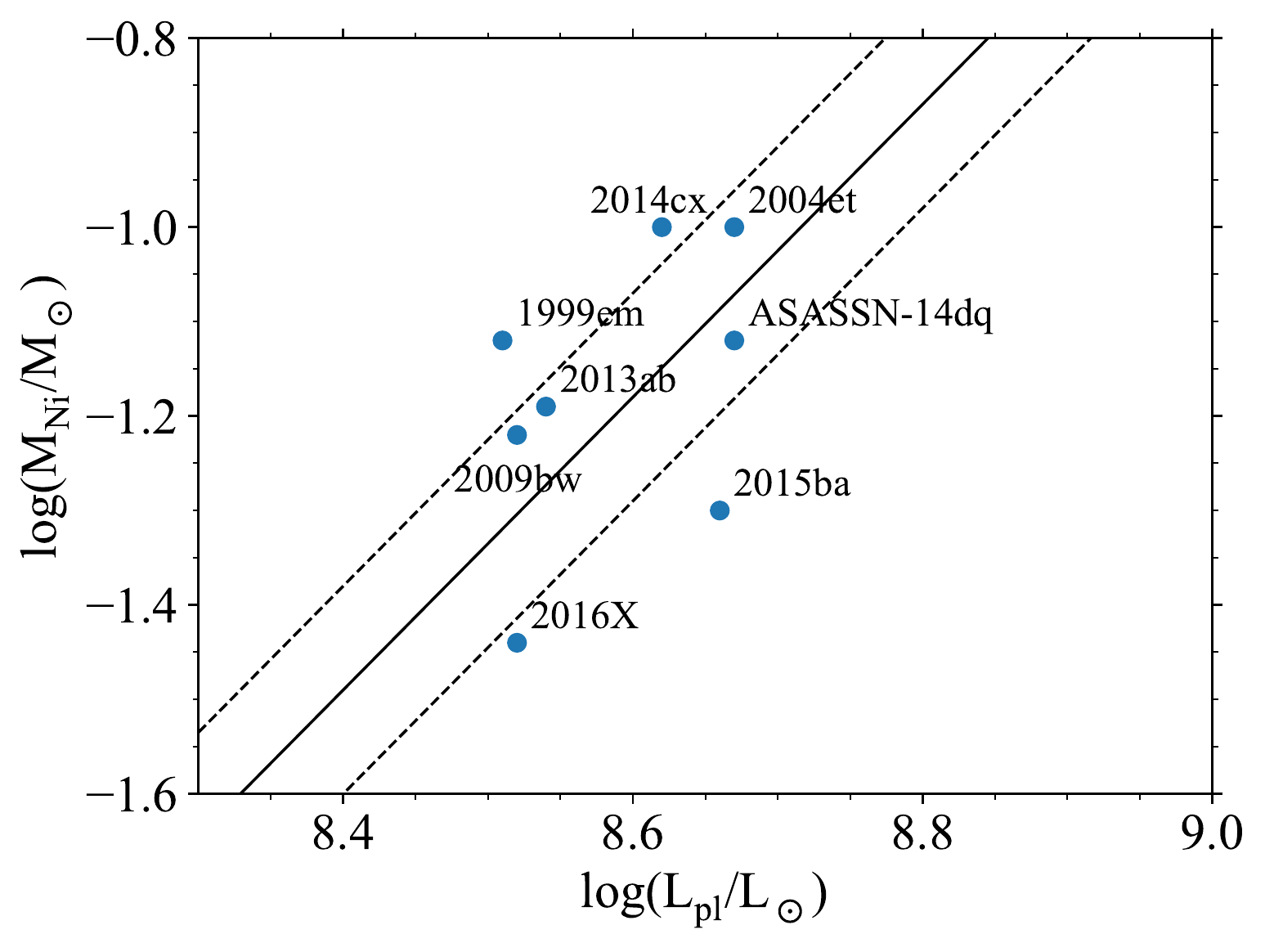}
\caption{The relationship between { $^{56}$Ni} mass and bolometric
  luminosity at 50 days for our models plotted against the correlation
  found by \citet{muller17}. The linear relation and its intrinsic
  width are shown as solid and dashed lines, respectively.
\label{fig:nimass}}
\end{figure}

Where $L_{pl}$ is the bolometric luminosity at 50 days post explosion. The values obtained from our fits are in reasonable agreement with
this relation (see Figure \ref{fig:nimass}), and are also in agreement
with relations found by \citet{pejcha15}.

\FloatBarrier

\begin{figure}[!htb] 
\includegraphics[width=\columnwidth]{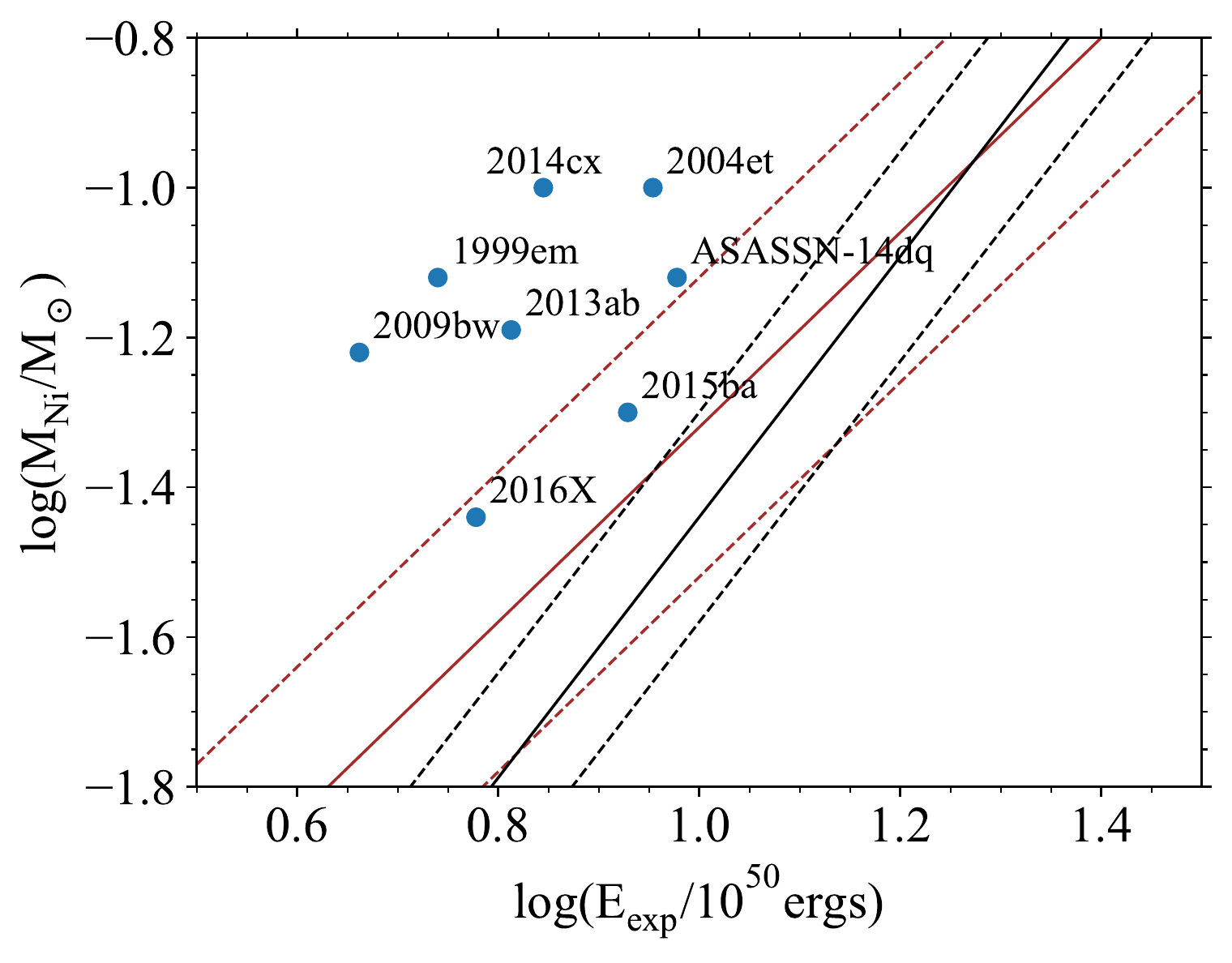}
\caption{The relationship between { $^{56}$Ni} mass and explosion
  energy for our models plotted against the correlations found by
  \citet{muller17} using the scaling relations of \citet{ln85} (black)
  and \citet{popov93} (brown). The linear relations and their
  intrinsic widths are shown as solid and dashed lines, respectively.
\label{fig:energy}}
\end{figure}

The explosion energies for the various SNe derived from our model fits
do not compare well with those found in previous work. No clear
pattern emerges between explosion energies derived from {\sc STELLA}
and those from either {\sc SNEC} or general-relativistic
radiation-hydrodynamical models. As noted though many of the fits
calculated with the latter seemed to underestimate the photospheric
velocities, and would therefore suggest a higher explosion energy.
Figure \ref{fig:energy} shows the comparison of our values with the
relation between { $^{56}$Ni} mass and explosion energy derived for
a large sample of SNe by \citet{muller17}, using the scaling relations
of \citet{ln85}:

\begin{equation}
\log(\frac{{ M_{^{56}Ni}}}{M_\odot})=1.74^{+0.30}_{-0.24}\,\log(\frac{E_{exp}}{10^{50}})-3.18^{+0.25}_{-0.23}
\end{equation}
 
A similar relation was also derived by \citet{muller17} using
  the scaling relations of \citet{popov93}:

\begin{equation}
\log(\frac{{ M_{^{56}Ni}}}{M_\odot})=1.30^{+0.28}_{-0.21}\,\log(\frac{E_{exp}}{10^{50}})-2.62^{+0.21}_{-0.16}
\end{equation}

Our values indicate a higher { $^{56}$Ni} mass for a given
explosion energy compared to both calibrations, with a large spread in
values. The systematic errors in our estimations are difficult to
quantify, given that they may depend on factors in the stellar
evolution, the explosion mechanism, or the mixing. A larger sample
size of SNe might perhaps provide a stronger correlation, but given
that all eight of our SNe fall above both \citet{muller17} relations,
this seems unlikely.  M18 show how the explosion energy can vary
depending on the degree of { $^{56}$Ni} mixing. \citet{pejcha15}
show that there is an inherent degeneracy between { $^{56}$Ni} mass
and explosion energy that makes the correlation weak. It is clear from
Figure 3 in \citet{muller17} that although there may be a correlation
between the $^{56}$Ni mass and explosion energy, the scatter in the
values is large.  Fig.~10 in M18 shows an equally large spread of
values.

\FloatBarrier

\section{Conclusions}

We have used the {\sc STELLA} code included in the newest release of
{\sc MESA} to determine the progenitor properties for a sample of
eight Type IIP SNe. We find that the version of {\sc STELLA} provided
is adequate to model both the lightcurves and Fe II 5169 velocities
for a wide range of SNe. It is able to provide reasonably
well-constrained SN parameters such as the progenitor ZAMS mass, total
explosion energy and synthesized { $^{56}$Ni} mass, among
others. In the past, hydrodynamical models have generally returned
progenitor ZAMS masses much higher than those derived by optical
progenitor identification, via X-ray or radio modelling, or late-time
spectral modelling. This was also true to some degree in the work of
M18. In our work we do not find this.  Of the eight SNe investigated
in this paper, we find that seven have progenitor ZAMS masses $\leq18$
$\msun$, with most in the 11-14 $\msun$ range. These results are in
agreement with past reports indicating that the majority of observed
Type IIPs have low mass progenitors. We do find one exception in SN
2015ba, whose lightcurve is clearly atypical compared to other Type
IIP SNe (Figure \ref{fig:lcs}). This SN requires a ZAMS mass likely
around 24 $\msun$, though our inability to accurately reproduce the
lightcurve of this particular SN introduces large uncertainties into
this result.  We note that this is not completely unexpected -
\citet{sukhboldetal16} claim that while most Type II SNe should arise
from below 20 $\msun$, $\sim 10\%$ of SNe arise from higher mass
stars.

The { $^{56}$Ni} masses in our study, although high, appear to fit
within the calibration of \citet{muller17} for the { $^{56}$Ni}
mass against the plateau luminosity at 50 days. This would perhaps
question the accuracy of { $^{56}$Ni} masses derived from
comparisons with the { $^{56}$Ni} value in SN 1987A.  However, when
plotted against the explosion energy following the relationships
derived by \citet{muller17}, we find that our { $^{56}$Ni} masses
appear high for the derived energy. Overall, while accepting that our
dataset is small, our values do not reflect tight relationships
between any of the parameters $^{56}$Ni mass, explosion energy,
progenitor mass and plateau luminosity. We agree with the results of
both \citet{pejcha15} and \citet{gutierrezetal17} which find that SN
explosions are not described by a single parameter but by a range of
parameter values.

\acknowledgements We thank the anonymous referee for a comprehensive
reading of the paper, and insightful comments and suggestions that
helped to improve the paper substantially. This work is supported by a
NASA Astrophysics Data Analayis program grant \#NNX14AR63G awarded to
PI V.~Dwarkadas at the University of Chicago. This work has made use
of the MESA stellar evolution code, and we are deeply grateful to the
authors.  We especially thank Bill Paxton and Jared Goldberg for
patiently answering all our questions about the way MESA works, and
Sergei Blinnikov for doing the same with the STELLA code. We
acknowledge useful discussions with Pablo Marchant and Mathieu Renzo.

\software{MESA \citep{mesa1,mesa2,mesa3,mesa4}, STELLA
  \citep{blinnikovetal98, bs04, blinnikovetal06}}




\bibliographystyle{aasjournal} \bibliography{iip}



\end{document}